\documentclass[10pt, conference, letterpaper]{IEEEtran}
\IEEEoverridecommandlockouts

\usepackage[plainruled,titlenumbered]{algorithm2e}
\makeatletter
\newcommand{\removelatexerror}{\let\@latex@error\@gobble}
\makeatother

\usepackage{amssymb}
\usepackage{amsmath}
\usepackage{amsfonts}       
\usepackage{mathtools}
\usepackage{nicefrac}       
\usepackage{amsthm}         
\usepackage{bbm}
\usepackage[helvet]{sfmath}
\usepackage[scaled]{helvet}
\allowdisplaybreaks

\usepackage{graphicx}
\graphicspath{ {figures/} }
\usepackage{epsfig}
\usepackage{subcaption}
\usepackage{float}

\usepackage[utf8]{inputenc} 
\usepackage[T1]{fontenc}
\usepackage{textcomp}

\usepackage{booktabs}       
\usepackage{makecell}

\usepackage{color}
\usepackage{xcolor}
\usepackage{xkeyval}

\usepackage{tikz}
\usetikzlibrary{bayesnet}
\usetikzlibrary{positioning}
\usetikzlibrary{arrows}
\usetikzlibrary{calc}

\usetikzlibrary{positioning,fit,shapes,backgrounds,circuits.logic.US}

\tikzstyle{server}=[circle, line width=0.5pt, rounded corners=0.1mm, draw=black!100, fill=tud3a!100]
\tikzstyle{vertex}=[circle, line width=1.5pt, draw=tud0d, fill=white]
\tikzstyle{dispatcher} =[and gate US, line width=0.5pt, draw=black!100, fill=tud0c!100]
\tikzstyle{dotbox} = [draw=white, fill=white, rectangle,  inner sep=10pt, inner ysep=20pt]

\tikzset{three_sided/.style={
		draw=none,rectangle, 
		append after command={
			[shorten <= -0.5\pgflinewidth]
			([shift={(-1.5\pgflinewidth,-0.5\pgflinewidth)}]\tikzlastnode.north west)
			edge([shift={( 0.5\pgflinewidth,-0.5\pgflinewidth)}]\tikzlastnode.north east)
			([shift={( 0.5\pgflinewidth,-0.5\pgflinewidth)}]\tikzlastnode.north east)
			edge([shift={( 0.5\pgflinewidth,+0.5\pgflinewidth)}]\tikzlastnode.south east)
			([shift={( 0.5\pgflinewidth,+0.5\pgflinewidth)}]\tikzlastnode.south east)
			edge([shift={(-1.0\pgflinewidth,+0.5\pgflinewidth)}]\tikzlastnode.south west)
		}
	}
}

\usepackage{paralist} 
\usepackage{enumitem}
\usepackage{amstext} 
\usepackage{array} 
\usepackage{url} 
\usepackage{xspace}
\usepackage{csquotes}
\usepackage{etoolbox}

\usepackage{cite}

\usepackage{ifthen}

\usepackage{balance}

\newboolean{longVersion}
\setboolean{longVersion}{true}  
\newcommand{\IfLongVersion}[1]{\ifthenelse{\boolean{longVersion}}{#1}{}}
\newcommand{\IfShortVersion}[1]{\ifthenelse{\boolean{longVersion}}{}{#1}}

\usepackage{todonotes}
\newboolean{todo}
\setboolean{todo}{true} 

\newcommand{\remarkInternal}[4]{\ifthenelse{\boolean{todo}}{\todo[inline, color=#2, caption={2do}, #3]{\begin{minipage}{\textwidth-4pt}\emph{Remark #1:}\\#4\end{minipage}}}{}}

\usepackage{blindtext}
\newboolean{blind}
\setboolean{blind}{false} 
\newcommand{\blindparagraph}[2][]{\ifthenelse{\boolean{blind}}{\blindtext[1]}{}}

\usepackage{hyphenat} 

\newcommand{\ie}{\textit{i.e.},\xspace}

\DeclareMathOperator*{\diag}{\mathrm{diag}\,}

\DeclareMathOperator*{\argmax}{\arg\max}
\DeclareMathOperator*{\E}{\mathrm{E}}
\DeclareMathOperator*{\GammaDis}{\mathrm{Gam}\,}

\DeclareMathOperator*{\NormalDis}{\mathcal{N}\,}
\DeclareMathOperator*{\GIGDis}{\mathcal{G}\mathcal{I}\mathcal{G}\,}

\title{CBA: Contextual Quality Adaptation for Adaptive Bitrate  Video Streaming \IfLongVersion{\\(Extended Version)}\thanks{This work has been funded by the German Research Foundation~(DFG) as part of the projects B4 and C3 within the Collaborative Research Center~(CRC) 1053 -- MAKI.}}

\author{\IEEEauthorblockN{Bastian Alt$^{+,}$\IEEEauthorrefmark{1},
\thanks{$^{+}$ The first two authors equally contributed major parts of this article.}
Trevor Ballard$^{+,}$\IEEEauthorrefmark{3},  Ralf Steinmetz\IEEEauthorrefmark{3}, Heinz Koeppl\IEEEauthorrefmark{1}, Amr Rizk\IEEEauthorrefmark{3}}

\IEEEauthorblockA{\IEEEauthorrefmark{1}Bioinspired Communication Systems Lab (BCS),  \{bastian.alt \textbar~heinz.koeppl\}@bcs.tu-darmstadt.de}
\IEEEauthorblockA{\IEEEauthorrefmark{3}Multimedia Communications Lab (KOM),  ballardt@knights.ucf.edu, \{amr.rizk \textbar~rst\}@kom.tu-darmstadt.de, \\
Technische Universit\"at Darmstadt, Germany}
}
\begin{document}

\maketitle
\begin{abstract}

Recent advances in quality adaptation algorithms leave adaptive bitrate (ABR) streaming architectures at a crossroads: When determining the sustainable video quality one may either rely on the information gathered at the client vantage point or on server and network assistance.
The fundamental problem here is to determine how valuable either information is for the adaptation decision.
This problem becomes particularly hard in future Internet settings such as Named Data Networking (NDN) where the notion of a network connection does not exist.

In this paper, we provide a fresh view on ABR quality adaptation for QoE maximization, which we formalize as a decision problem under uncertainty, and for which we contribute a sparse Bayesian contextual bandit algorithm denoted \emph{CBA}.
This allows taking high-dimensional streaming context information, including client-measured variables and network assistance, to find \emph{online} the most valuable information for the quality adaptation.
Since sparse Bayesian estimation is computationally expensive, we develop a fast new inference scheme to support online video adaptation.
We perform an extensive evaluation of our adaptation algorithm in the particularly challenging setting of NDN, where we use an emulation testbed to demonstrate the efficacy of CBA compared to state-of-the-art algorithms.

\end{abstract}


\section{Introduction}\label{sec:intro}

Video streaming services such as Netflix, YouTube, and Twitch, which constitute an overwhelming share of current Internet traffic, use adaptive bitrate streaming algorithms that try to find the most suitable video quality representation given the client's networking conditions. Current architectures use Dynamic Adaptive Streaming over HTTP (DASH) in conjunction with client-driven algorithms to adjust the quality bitrate of each video segment based on various signals, such as measured throughput, buffer filling, and derivatives thereof. In contrast, new architectures such as SAND~\cite{sand} 
introduce network-assisted streaming via DASH-enabled network elements that provide the client with guidance, such as accurate throughput measurements and source recommendations. Given the various adaptation algorithms that exist in addition to client-side and network-assisted information, a fundamental question arises on the importance of this context information for the Quality of Experience (QoE) of the video stream.

The problem of video quality adaptation is aggravated in Future Internet architectures such as Named Data Networking (NDN). In NDN, content is requested by name rather than location, and each node within the network will either return the requested content or forward the request.
Routers are equipped with caches to hold frequently-requested content, thereby reducing the round-trip-time (RTT) of the request while simultaneously saving other network links from redundant content requests.
Several attempts to make DASH-style streaming possible over NDN exist, e.g., \cite{samain2017}, for which the key difficulty is that traditional algorithms rarely play to the strengths of NDN where the notion of a connection does not exist.
Throughput, for example, is not a trivial signal in NDN as data may not be coming from the same source.

\begin{figure}
	\includegraphics[width=\columnwidth]{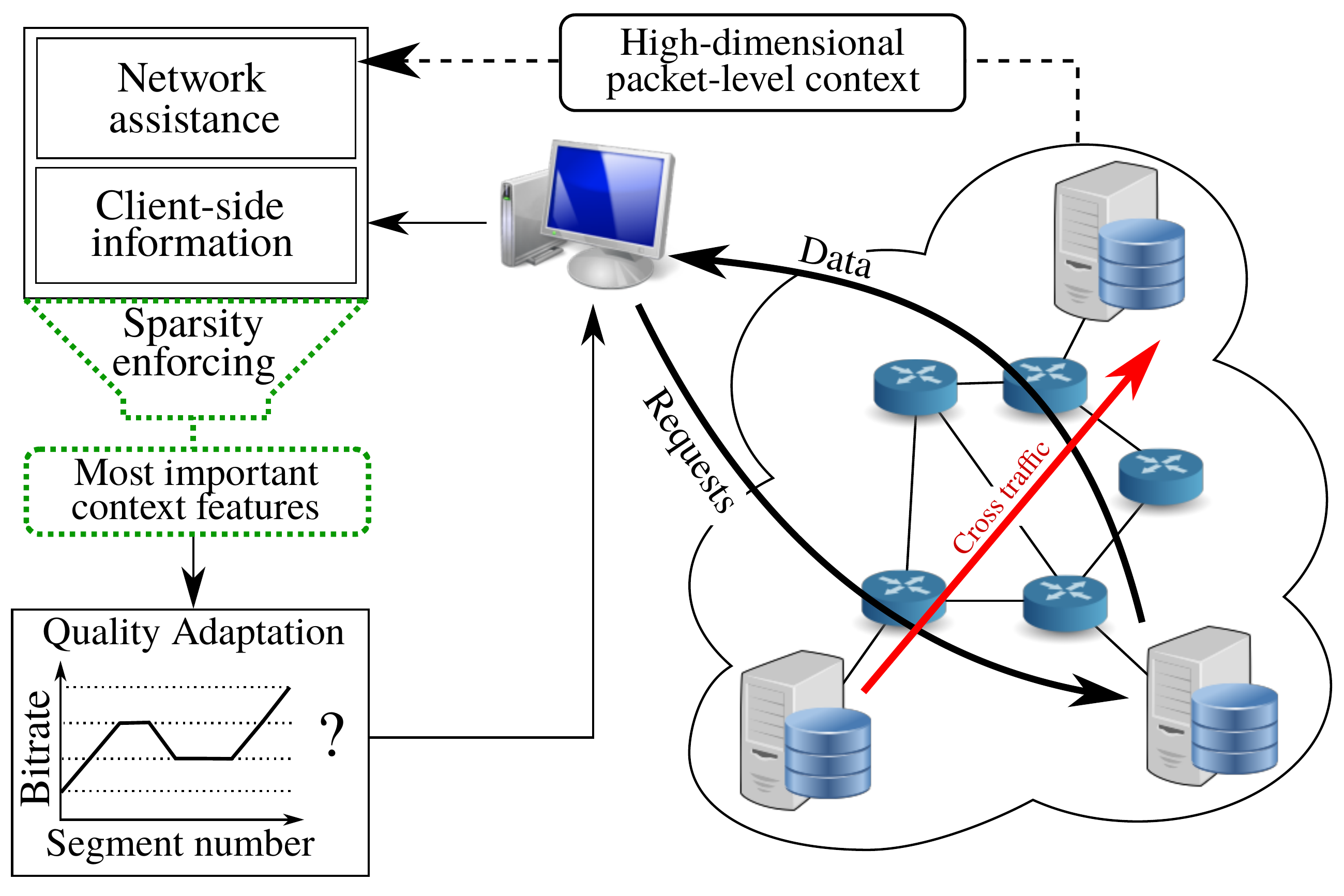}
	\caption{A standard client-based and/or network-assisted ABR streaming model (black) with the proposed Context-based Adaptation---CBA (dotted). In CBA, high-dimensional context features from the network, along with client-side information, undergo sparsity enforcement to shrink the impact of unimportant features.}
	\label{fig:net_model}
\end{figure}

In this paper, we closely look at the problem of using context information available to the client for video quality adaptation.
Note that our problem description is agnostic to the underlying networking paradigm, making it a good fit to traditional IP-based video streaming as well as NDN.
In essence, we consider the fundamental problem of sequential decision-making under uncertainty where the client uses network context information received with every fetched video segment.
In Fig.~\ref{fig:net_model} we show a sketch where the client adaptation algorithm decides on the quality of the next segment based on a high-dimensional network context.
We model the client's decision on a video segment quality as a contextual multi-armed bandit problem aiming to optimize an objective QoE metric that comprises \emph{(i)} the average video quality bitrate, \emph{(ii)} the quality degradation, and \emph{(iii)} the video stalling.

%


One major challenge with incorporating high-dimensional network context information in video quality adaptation is extracting the information that is most relevant to the sought QoE metric.
We note that the interactions within this context space become complicated given the NDN architecture, where the network topology and cache states influence the streaming session.
Our approach introduces a sparse Bayesian contextual bandit algorithm that is fast enough to run online during video playback.
The rationale behind the sparsity is that the given information, including network-assisted and client-side measured signals such as buffer filling and throughput, constitutes a high-dimensional context which is difficult to model in detail.
Our intuition is that, depending on the client's network context, only a few input variables have a significant impact on QoE.
Note, however, that sparse Bayesian estimation is usually computationally expensive.
Hence, we develop here a fast new inference scheme to support online quality adaptation.

Our contributions in this paper can be summarized as:

\begin{itemize}
  \item We formulate the quality adaptation decision for QoE maximization in ABR video streaming as a contextual multi-armed bandit problem.
  \item We provide a sparse Bayesian contextual bandit algorithm, denoted \emph{CBA}, which is computationally fast enough to provide real-world video players with quality adaptation decisions based on the network context.
  \item We show emulation testbed results and demonstrate the fundamental differences to the established state-of-the-art quality adaptation algorithms, especially given an NDN architecture.
\end{itemize}

The developed software is provided here\footnote{https://github.com/arizk/cba-pipeline-public}. The remainder of this paper is organized as follows: In Sect.~\ref{sec:related_work}, we review relevant related work on ABR video streaming and contextual bandits. In Sect.~\ref{sec:system_model}, we present the relevant background on ABR video streaming. In Sect.~\ref{sec:multi_armed_bandits}, we model the quality adaptation problem as a contextual multi-armed bandit problem before providing a fast contextual bandit algorithm for high-dimensional information. In Sect.~\ref{sec:video_adaptation_as_CB}, we show how ABR streaming uses CBA and define a QoE-based reward. We describe the evaluation testbed before providing emulation results in Sect.~\ref{sec:evaluation}. Section~\ref{sec:conclusion} concludes the paper.

\section{Related Work}
\label{sec:related_work}
In the following, we split the state-of-the-art related work into two categories; i.e., work on ABR quality adaptation, especially in NDN, and related work on contextual bandit algorithms with high-dimensional covariates.

Significant amounts of research have been given to finding streaming architectures capable of satisfying high bitrate and minimal rebuffering requirements at scale. CDN brokers such as Conviva \cite{conviva} allow content producers to easily use multiple CDNs, and are becoming crucial to meet user demand \cite{mukerjee2016}. Furthermore, the use of network assistance in CDNs has received significant attention recently as a method of directly providing network details to DASH players. SAND \cite{sand} is an ISO standard which permits DASH enabled in-network entities to communicate with clients and offer them QoS information.
SDNDASH \cite{bentaleb2016} is another such architecture aiming to maintain QoE stability across clients, as clients without network assistance information are prone to misjudge current network conditions, causing QoE to oscillate. Beyond HTTP, the capabilities of promising new network paradigms such as NDN pose challenges to video streaming. The authors of  \cite{samain2017} compare three state-of-the-art DASH adaptation algorithms over NDN and TCP/IP, finding NDN performance to notably exceed that of TCP/IP given certain network conditions. New adaptation algorithms specific to NDN have also been proposed, such as NDNLive \cite{wang2016}, which uses a simple RTT mechanism to stream live content with minimal rebuffering.

In this work, we model the video quality adaptation problem as a contextual bandit problem assuming a linear parametrization, which has successfully been used, e.g., for ad placement \cite{li2010}. Another promising approach is based on cost-sensitive classification in the bandit setting \cite{agarwal2014}. Recently, \cite{urteaga18} has discussed the use of variational inference in the bandit setting, wherein Thompson sampling is considered to cope with the exploration-exploitation trade-off.
By assuming a high-dimensional linear parametrization, we make use of sparse estimation techniques. High-dimensional information arises in video streaming due to the network context. Sparsity has been a major topic in statistical modeling and many Bayesian approaches have been proposed. Traditionally, double exponential priors which correspond to $\ell_1$ regularization have been used. However, these priors often fail due to limited flexibility in their shrinkage behavior. Other approaches that induce sparsity include 'spike-and-slab' priors \cite{ishwaran2005} and continuous shrinkage priors. Between these two, continuous shrinkage priors have the benefit of often being computationally faster \cite{armagan2011}.
For our approach we use the Three Parameter Beta Normal (TPBN) continuous shrinkage prior introduced by \cite{armagan2011}, which generalizes diverse shrinkage priors, e.g, the horseshoe prior \cite{carvalho2009}, the Strawderman-Berger prior,
the normal-exponential-gamma prior,
and the normal-gamma prior.

\section{Adaptive Bitrate Streaming: Decisions under Uncertainty}
\label{sec:system_model}
In this section, we review the established model for quality adaptation in ABR video streaming and highlight the changes that arise when streaming over NDN.


\subsection{Adaptive Bitrate Streaming: A Primer}

In adaptive bitrate streaming, the content provider offers multiple qualities of the same content to clients, who decide which one to pick according to their own client-side logic.
Each video
is divided into $T$ consecutive segments which represents some fixed $L$ seconds of content. These segments are encoded at multiple bitrates corresponding to the perceived average segment quality.
In practice, segment lengths are often chosen to be two to ten seconds~\cite{Stohr:2017} with several distinct quality levels to choose from, such as 720p and 1080p.
Let $\mathcal{V}$ represent the set of all available video qualities, such that $\mathcal{V}=\{v(1), v(2), ..., v(K)\}$ and $v(i) > v(j)$ for all $i > j$; i.e., a higher index indicates a higher bitrate and better quality.
Let the $t$-th segment encoded at the $i$-th quality be denoted $s_{t}(i)$.

Received video segments are placed into a playback buffer which contains downloaded, unplayed video segments.
Let the number of seconds in the buffer when segment $t$ is received be $B_t$, and let the playback buffer size \texttt{BUF\_MAX} be the maximum allowed seconds of video in the buffer.
By convention, we define $B_0=0$ and write the recursion of the buffer filling as $B_t = \max\{B_{t-1} + L - \xi_{t}(i),L\}$, where $\xi_{t}(i)$ denotes the fetch time for $s_{t}(i)$.
A stalling event is ascribed to the $t$-th segment when $\xi_{t}(i) > B_{t-1}$.
Note that the recursion above holds only if $B_{t-1} + L < \texttt{BUF\_MAX}$; i.e., the client is blocked from fetching new segments if the playback buffer is full. If this occurs, the client idles for exactly $L$ seconds before resuming segment fetching.
In some related work~\cite{Stohr:2017}, \texttt{BUF\_MAX} is chosen between 10 and 30 seconds.

To allow the client to select a segment in the $i$-th quality, the client fetches a Media Presentation Description (MPD), an XML-like file with information on the available video segments and quality levels, during session initialization.
After obtaining the MPD, the client may begin to request each segment according to its adaptation algorithm.
In general, uncertainty exists over the segment fetch time.
The most prevalent quality adaptation algorithms take throughput estimates \cite{Panda14} or the current buffer filling $B_t$ \cite{Bola16}, or combinations and functions thereof to make a decision on the quality of the next segment $s_{t+1}(i)$.
The decision aims to find the segment quality which maximizes a QoE metric, such as the average video bitrate, or compound metrics taking the bitrate, bitrate variations, and stalling events into account.

%

%

\subsection{Streaming over Named Data Networking}

In NDN, consumers or clients issue interests which are forwarded to content producers, i.e., origin servers, via caching-enabled network routers. These interests are eventually answered with data provided by the producer or an intermediary router cache. To request a video, a consumer will first issue an interest for the MPD of the video. Each $s_{t}(i)$ is given a name in the MPD, e.g., of the form \texttt{/video ID/quality level/segment number}.
The client issues an interest for each data packet when requesting a particular segment.
Since NDN data packets are of a small, fixed size, higher-quality video segments will require more data packets to encode. We do not permit the client to drop frames, so all data packets belonging to some segment $s_{t}(i)$ must be in the playback buffer to watch that segment.

\section{A Fast Contextual Bandit Algorithm for High Dimensional Covariates}
\label{sec:multi_armed_bandits}
In this work, we model the problem of video quality adaptation as a sequential decision-making problem 
under uncertainty, for which a successful framework is given by the multi-armed bandit problem dating back to~\cite{robbins1952}. The contextual bandit problem \cite{bubeck2012} is an extension to the classic problem, where additional information is revealed sequentially. The decision-making can therefore be seen as a sequential game.

At decision step $t$, i.e., at the $t$-th segment, a learner observes a $D$ dimensional context variable $\mathbf{x}_{t}(a) \in \mathbb{R}^D$ for a set of $K$ actions $a \in \{1,\dots,K\}$. Here, the actions map to the $K$ video qualities that the client chooses from. The client chooses an action $a_t$, for which it observes a reward $r_{t}(a_t)$. This reward can be measured in terms of low-level metrics such as fetching time or, as we consider later, QoE. The decision making is performed over a typically unknown decision horizon $T$, \ie $t \in \{1, \dots, T\}$. Therefore, the learner tries to maximize the cumulative reward $\sum_{t=1}^T r_{t}(a_t)$ until the decision horizon. It is important to note that after each decision the learner only observes the reward $r_{t}(a_t)$ associated to the played action $a_t$; hypothetical rewards for other actions $a\neq a_t$, are not revealed to the learner.

Next, we model the contextual bandit problem under the linearizability assumption, as introduced in \cite{auer2002}. Here, we assume that a parameter $\pmb{\beta}^\ast(a) \in \mathbb{R}^D$  controls the mean reward of each action $a$ at decision step $t$ as $\E[r_{t}(a)]={\mathbf{x}_{t}(a)}^\top \pmb{\beta}^{\ast}(a)$.  We introduce the regret $R_T$ of an algorithm to evaluate its performance as
\begin{equation}
R_T=\sum_{t=1}^T r_{t}(a_t^\ast)-\sum_{t=1}^T r_{t}(a_t),
\label{eq:regret}
\end{equation}
with $a_t^\ast=\argmax_{a} {\mathbf{x}_{t}(a)}^\top \pmb{\beta}^{\ast}(a)$. The regret compares the cumulative reward of the algorithm against the cumulative reward with hindsight.
In order to develop algorithms with a small regret in the linear setting, many different strategies have been proposed. Such algorithms include techniques based on forced sampling \cite{bastani2015}, Thompson sampling \cite{agrawal2013}, and the upper confidence bound (UCB) \cite{auer2002,li2010,chu2011,krause2011}.

Network-assisted video streaming environments provide high-dimensional context information, so it is natural to assume a sparse parameter $\pmb{\beta}^{\ast}(a)$. We therefore impose a sparsity-inducing prior on the sought regression coefficients $\pmb{\beta}(a)$. To cope with the contextual bandit setting, we start with the Bayes-UCB algorithm with liner bandits introduced in \cite{kaufmann2012} and develop a version which fits the given problem. Since previously developed sparse Bayesian inference algorithms are computationally expensive, we develop a fast new inference scheme for the contextual bandit setting.

\subsection{The Contextual Bayes-UCB Algorithm - CBA}
\label{sec:cba_ucb}
The Contextual Bayes-UCB algorithm (CBA-UCB) selects in each round the action $a$ which maximizes the index
\begin{equation}
q_t(a)=Q(1-\frac{1}{\alpha t}, \zeta_{t-1}(a)),
\label{eq:idx_general}
\end{equation}
where $\alpha$ is a width parameter for the UCB and $Q(t, \rho)$ is the quantile function associated with the distribution $\rho$, \ie $\mathbb{P}(X \leq Q(t, \rho))=t$, with $X \sim \rho$. Additionally, we denote $\zeta_{t-1}(a)$ as the posterior distribution  of the mean reward
\begin{equation}
\zeta_{t-1}(a)=p({\mathbf{x}_{t}(a)}^\top \pmb{\beta}(a) \mid \mathcal{D}_{t-1}(a)),
\label{eq:posterior_mean_reward}
\end{equation}
where $\mathcal{D}_{t-1}(a)$ is the set of data points of contexts and rewards for which action $a$ was previously played
\begin{equation}
\mathcal{D}_{t-1}(a)=\{(\mathbf{x}_{t^\prime}(a), r_{t^\prime}(a),a_{t^\prime}): a_{t^\prime}=a, 1\leq t^\prime \leq t-1\}.
\end{equation}
In the following subsections, we derive a Gaussian distribution for the posterior distribution  of the regression coefficients $p(\pmb{\beta}(a)\mid \mathcal{D}_{t-1}(a))=\NormalDis(\pmb{\beta}(a) \mid \pmb{\mu}_{\pmb{\beta}(a)}, \pmb{\Sigma}_{\pmb{\beta}(a)})$.
In this case the index in \eqref{eq:idx_general} reduces to
\begin{equation}
q_t(a)={\mathbf{x}_{t}(a)}^\top \pmb{\mu}_{\pmb{\beta}}(a)+ Q\left(1-\frac{1}{\alpha t}, \NormalDis(0,1)\right)\sqrt{{\mathbf{x}_{t}(a)}^\top \pmb{\Sigma}_{\pmb{\beta}(a)}\mathbf{x}_{t}(a)},
\label{eq:idx_gaussian}
\end{equation}
where the quantile function
computes to $\sqrt{2}\mathrm{erf}^{-1}(1-\frac{2}{ \alpha  t})$, with the inverse error function $\mathrm{erf}^{-1}(\cdot)$. \IfLongVersion{The algorithm for CBA-UCB is depicted in Fig.~\ref{fig:algorithms}, Alg.~1.}

\IfLongVersion{
\begin{figure*}
  \removelatexerror
\begin{subfigure}[t]{\columnwidth}
\centering
\begin{algorithm}[H]
\TitleOfAlgo{{\sc main-routine} CBA-UCB with Gaussian Posterior}
\input{algorithms/algorithm_main}
\end{algorithm}
\end{subfigure}%
~
\begin{subfigure}[t]{\columnwidth}
\centering
\begin{algorithm}[H]
\TitleOfAlgo{{\sc sub-routine} SVI}
\input{algorithms/algorithm_svi}
\end{algorithm}
\end{subfigure}

\begin{subfigure}[t]{\columnwidth}
\centering
\begin{algorithm}[H]
\TitleOfAlgo{{\sc sub-routine} VB}
\input{algorithms/algorithm_vb}
\end{algorithm}
\end{subfigure}
~
\begin{subfigure}[t]{\columnwidth}
\centering
\begin{algorithm}[H]
\TitleOfAlgo{{\sc sub-routine} OS-SVI}
\input{algorithms/algorithm_ossvi}
\end{algorithm}
\end{subfigure}
\caption{The CBA-UCB Algorithm with three Bayesian inference schemes for the regression coefficients: Variational Bayesian Inference (VB), Stochastic Variational Inference (SVI) and  One Step Stochastic Variational Inference (OS-SVI). }
\label{fig:algorithms}
\end{figure*}
}

\subsection{Generative model of the linear rewards}
\label{subsec:generative_model}
Here, we derive the posterior inference for the regression coefficients $\pmb{\beta}(a)$. The posterior distributions are calculated for each of the $K$ actions. For the inference of the posterior \eqref{eq:posterior_mean_reward}, we use Bayesian regression to infer the posterior of the regression coefficients\footnote{For readability we drop the dependency on $a$ of the regression coefficients $\pmb{\beta}(a)$} $\pmb{\beta}=[\beta_1,\dots, \beta_D]^\top$.  We use the data $\mathcal{D}_{t-1}(a)$, which is a set of $M$ previously observed contexts $\mathbf{X}=[\mathbf{x}_1,\dots,\mathbf{x}_{M}]^\top$ and rewards $\mathbf{r}=[r_1,\dots,r_{M}]^\top$ when taking action $a$.

Assuming a linear regression model $\mathbf{r}=\mathbf{X}\pmb{\beta}+\pmb{\epsilon},$ with i.i.d. noise $\pmb{\epsilon} \sim \NormalDis(\pmb{\epsilon} \mid \mathbf{0}, \mathbf{I}_M/\sigma^{-2})$ the regression response $\mathbf{r}$ follows the likelihood
$$p(\mathbf{r} \mid \pmb{\beta}, \sigma^{-2})=\prod_{m=1}^M p(r_m \mid \pmb{\beta}, \sigma^{-2}) =\prod_{m=1}^M  \NormalDis(r_m \mid\mathbf{x}_m^\top \pmb{\beta}, 1/\sigma^{-2}),$$
where $\sigma^{-2}$ is the noise precision for the regression problem.
For the application of video streaming with high-dimensional context information, we use a sparsity inducing prior over the regression coefficients $\pmb{\beta}$ to find the most valuable context information.
We use here the Three Parameter Beta Normal (TPBN) continuous shrinkage prior introduced by \cite{armagan2011}
, which puts on each regression coefficient $\beta_j$, $j \in \{1, \dots, D\}$, the following hierarchical prior
\begin{equation}
\begin{split}
&\beta_j \sim \NormalDis(\beta_j\mid0,\tau_j/\sigma^{-2}), \; \tau_j \sim \GammaDis(\tau_j\mid a_0, \lambda_j),\\
&\lambda_j \sim \GammaDis(\lambda_j\mid b_0,\phi),
\label{eq:tpbn_prior}
\end{split}
\end{equation}
where $\tau_j$ is a Gamma distributed\footnote{We use the shape and rate parametrization of the Gamma distribution.} continuous shrinkage parameter that shrinks $\beta_j$, as $\tau_j$ gets small.
The parameter $\lambda_j$ controls $\tau_j$ via a global shrinkage parameter parameter $\phi$. For appropriate hyper-parameter choice of $a_0$ and $b_0$ different shrinkage prior are obtained. For example we use $a_0=1/2$, $b_0=1/2$ which corresponds to the horseshoe prior~\cite{carvalho2009}. For notational simplicity, we collect the parameters $\lambda_j, \tau_j$ for the context dimensions $j\in\{1, \dots, D\}$ in the column vectors $\pmb{\lambda},\pmb{\tau}$, respectively.

\begin{figure}
  \begin{center}
  	\resizebox {0.85\columnwidth} {!} {
      \begin{tikzpicture}

  
  \node[obs]	(r){$r_m$};
  \factor[above=of r] {r_f} {right:$\NormalDis$} {} {} ; %
  \node[det, above left= of r_f] (dot_b_x) {dot};
  \node[const, above right= of dot_b_x] (x_m){$\mathbf{x}_m$};
  \node[latent, left=2cm of r ] (beta_j){$\beta_j$};
  \factor[above = of beta_j]{beta_f} {right:$\NormalDis$} {} {} ; %
  \node[const, above left=2.5cm of beta_f](zero) {$0$};
  \node[det,  above left = .2cm and 2.35cm of x_m](dot_t_b){dot};
  \node[latent, above left=of dot_t_b](tau_j){$\tau_j$};
   \node[det, right= 3cm of dot_t_b] (inv) {$(\cdot)^{-1}$};
    \node[latent, above = of inv](sigma){$\sigma^{-2}$};
  \factor[above = of sigma]{sigma_f}{right:$\GammaDis$}{}{};
  \node[const, above left=of sigma_f](c){$c_0/2$};
  \node[const, above right= of sigma_f](d){$d_0/2$};
  \factor[above = of tau_j]{tau_f}{right:$\GammaDis$}{}{};
  \node[const, above left = of tau_f] (a) {$a_0$};
  \node[latent, above right = of tau_f](lambda_j){$\lambda_j$};
  \factor[above = of lambda_j]{lambda_f}{above:$\quad\GammaDis$}{}{};
  \node[const, above left = .7cm and .5cm of lambda_f](b){$b_0$};
  
  \node[latent,  right = 2.5cm of tau_j](phi){$\phi$};
   \factor[above = of phi]{phi_f}{right:$\GammaDis$}{}{};
   \node[const, above left =.8cm and .5cm  of phi_f](half){$1/2$};
   \node[latent, above  = of phi_f](omega){$\omega$};
    \factor[above = of omega]{omega_f}{right:$\GammaDis$}{}{};
      \node[const, above left = of omega_f](halfa){$1/2$};
        \node[const, above right = of omega_f](halfb){$1$};
        
 \factoredge {dot_b_x,inv} {r_f} {r} ; %
  \edge[-] {beta_j,x_m} {dot_b_x} ;
   \edge[-] {sigma} {inv} ;
    \factoredge {zero,dot_t_b} {beta_f} {beta_j} ; %
      \edge[-] {tau_j,inv} {dot_t_b} ;
      \factoredge {a,lambda_j} {tau_f} {tau_j} ; %
       \factoredge {c,d} {sigma_f} {sigma} ; %
        \factoredge {b,phi} {lambda_f} {lambda_j} ; %
         \factoredge {half,omega} {phi_f} {phi} ; %
            \factoredge {halfa,halfb} {omega_f} {omega} ; %
            
   \plate {rx} { %
    (r)(r_f)(r_f-caption) %
    (x_m) %
    (dot_b_x) %
  } {$m=1, \dots, M$} ;
     \plate {} { %
    (beta_j)(beta_f)(beta_f-caption) %
    (dot_t_b)
    (tau_j)(tau_f)(tau_f-caption) %
    (lambda_j)(lambda_f)(lambda_f-caption) %
  } {$j= 1, \dots, D$} ;

\end{tikzpicture}
    }
  \end{center}
  \caption{Probabilistic graphical model for the Bayesian regression in Sect.~\ref{subsec:generative_model} with Three Parameter Beta Normal prior using factor graph notation. (Deterministic functions are depicted in diamond-shaped nodes and 'dot' denotes the inner product.)}
  \label{fig:pgm}
\end{figure}
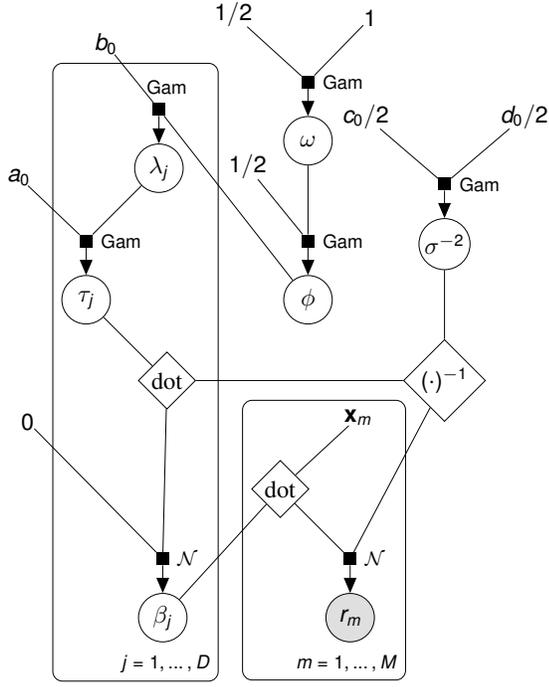

For the estimation of the global shrinkage parameter $\phi$ an additional hierarchy is used as $\phi \sim \GammaDis(\phi \mid 1/2, \omega)$ and  $\omega \sim \GammaDis(\omega \mid 1/2, 1)$. For the noise precision a gamma prior  is used $\sigma^{-2} \sim \GammaDis(\sigma^{-2}\mid c_0/2,d_0/2)$, with hyper parameters $c_0$ and $d_0$.
The graphical model \cite{dietz2010} of this generative model is depicted in Fig.~\ref{fig:pgm}.

\subsection{Variational Bayesian Inference (VB)}
\label{subsec:variational_bayes_inference}
In the following, we review the general approximate inference scheme of mean field variational Bayes (VB) and the application to the linear regression with TPBN prior as proposed in \cite{armagan2011}.
Thereafter, we leverage stochastic variational inference (SVI) to develop a new contextual bandit algorithm.

Since exact inference of the posterior distribution $p(\pmb{\beta},\\ \sigma^{-2}, \pmb{\tau}, \pmb{\lambda}, \phi, \omega \mid \mathbf{r})$ is intractable \cite{bishop2006}, we apply approximate inference in form of variational Bayes (VB) for posterior inference. We use
a mean field variational approximation, with $q(\pmb{\beta}, \sigma^{-2}, \pmb{\tau},\pmb{\lambda}, \phi, \omega)=q(\pmb{\beta})q(\sigma^{-2})q( \pmb{\tau})q(\pmb{\lambda})q(\phi)q(\omega)$ for the approximate distribution. The variational distributions are obtained by minimizing the Kullback-Leibler (KL) divergence between the variational distribution and the intractable posterior distribution
\begin{equation}
\begin{split}
\mathrm{KL}&(q(\pmb{\beta}, \sigma^{-2}, \pmb{\tau},\pmb{\lambda}, \phi, \omega)\Vert p(\pmb{\beta}, \sigma^{-2}, \pmb{\tau},\pmb{\lambda}, \phi, \omega \mid \mathbf{r}))\\
=&\mathrm{E}_q[\log(q(\pmb{\beta}, \sigma^{-2}, \pmb{\tau},\pmb{\lambda}, \phi, \omega))]\\
&-\mathrm{E}_q[\log(p(\pmb{\beta}, \sigma^{-2}, \pmb{\tau},\pmb{\lambda}, \phi, \omega , \mathbf{r}))]\\
&\quad+\log(p(\mathbf{r})).
\end{split}
\label{eq:kl_divergence}
\end{equation}
By Jensen's inequality, a lower bound on the marginal likelihood (evidence) can be found
\begin{equation}
\begin{split}
\mathcal{L}(q) &=\mathrm{E}_q[\log(p(\pmb{\beta}, \sigma^{-2}, \pmb{\tau},\pmb{\lambda}, \phi, \omega , \mathbf{r}))]\\
&\quad-\mathrm{E}_q[\log(q(\pmb{\beta}, \sigma^{-2}, \pmb{\tau},\pmb{\lambda}, \phi, \omega))]\\
&\leq \log(p(\mathbf{r})).
\end{split}
\label{eq:elbo_general}
\end{equation}
The evidence lower bound (ELBO) $\mathcal{L}(q)$ is used for solving the optimization problem over the KL divergence \eqref{eq:kl_divergence}, since maximizing $\mathcal{L}(q)$ is equivalent to minimizing the KL divergence. Using calculus of variations \cite{bishop2006}, the solution of the optimization problem can be found with the following optimal variational distributions\footnote{$\GIGDis(x\mid p,a,b)$ denotes the generalized inverse Gaussian distribution\IfLongVersion{, see Appendix \ref{sec:appendix_gig}}.}
\begin{equation}
\begin{split}
&q(\pmb{\beta})=\NormalDis(\pmb{\beta} \mid \pmb{\mu}_{\pmb{\beta}}, \pmb{\Sigma}_{\pmb{\beta}}),\;
q( \sigma^{-2})= \GammaDis(\sigma^{-2} \mid c^\ast, d^\ast),\; \\
&q(\pmb{\tau}) = \prod_{j=1}^{D} \GIGDis(\tau_j \mid p_{\tau, j}, a_{\tau, j}, b_{\tau, j}),\;\\
 &q(\pmb{\lambda})=\prod_{j=1}^{D} \GammaDis(\lambda_j \mid a_{\lambda,j},b_{\lambda,j}),\;
 q(\phi)= \GammaDis(\phi \mid a_{\phi}, b_{\phi}),\; \\
 &q(\omega)= \GammaDis(\omega \mid a_{\omega},b_{\omega}),
\end{split}
\label{eq:variational_distr}
\end{equation}
\IfLongVersion{with the parameters of the variational distributions
\begin{equation}
\begin{split}
&\pmb{\mu}_{\pmb{\beta}}=(\mathbf{X}^\top \mathbf{X}+ \mathbf{T}^{-1})^{-1} \mathbf{X}^\top \mathbf{r}\\
&\pmb{\Sigma}_{\pmb{\beta}}=\langle \sigma^{-2}\rangle^{-1}  (\mathbf{X}^\top \mathbf{X}+ \mathbf{T}^{-1})^{-1},\\
&\mathbf{T}^{-1}=\diag(\langle \tau_1^{-1}\rangle,\dots,\langle \tau_D^{-1}\rangle),\;
   c^{\ast}=\frac{M+D+c_0}{2},\;\\
    &d^{\ast}=(\mathbf{r}^\top \mathbf{r} - 2\mathbf{r}^\top\mathbf{X}\langle \pmb{\beta}\rangle + \sum_{m=1}^M \mathbf{x}_m^\top \langle \pmb{\beta}\pmb{\beta}^{\top}\rangle \mathbf{x}_m  \\
    &\qquad +\sum_{j=1}^D \langle \beta_j^2 \rangle \langle \tau_j^{-1} \rangle+d_0)/2,\;\\
    &p_{\tau, j}=a_0-1/2,\;
a_{\tau, j}=2 \langle \lambda_j \rangle,\;
b_{\tau, j}=\langle \beta_j^2 \rangle \langle \sigma^{-2} \rangle,\;\\
&a_{\lambda,j}=a_0+b_0,\;
b_{\lambda,j}= \langle \tau_j \rangle + \langle \phi \rangle,\;\\
&a_{\phi}=Db_0+1/2,\;
b_{\phi}= \langle \omega \rangle + \sum_{j=1}^D \langle \lambda_j \rangle,\;\\
&a_{\omega}=1,\;
b_{\omega}= \langle \phi \rangle +1
\end{split}
\label{eq:variational_param}
\end{equation}
and the moments
\begin{equation}
\begin{split}
&\langle \pmb{\beta}\rangle=\pmb{\mu}_{\pmb{\beta}},\;
\langle \pmb{\beta} \pmb{\beta}^\top\rangle= \pmb{\Sigma}_{\pmb{\beta}}+\pmb{\mu}_{\pmb{\beta}}\pmb{\mu}_{\pmb{\beta}}^\top,\;\\
&\langle \sigma^{-2} \rangle = c^\ast / d^\ast,\;
\langle \lambda_j \rangle= a_{\lambda,j}/b_{\lambda,j},\;
\langle \phi \rangle = a_{\phi}/ b_{\phi},\;\\
&\langle \omega \rangle = a_{\omega}/b_{\omega},\;
v_j=\sqrt{2 a_{\tau, j} b_{\tau, j}},\;\\
&\langle \tau_j \rangle = \left(\frac{b_{\tau, j}}{a_{\tau, j}}\right)^{1/2} \frac{\mathrm{K}_{p_{\tau, j}+1}(v_j)}{\mathrm{K}_{p_{\tau, j}}(v_j)},\;\\
&\langle \tau_j^{-1} \rangle =\left(\frac{a_{\tau, j}}{b_{\tau, j}}\right)^{1/2} \frac{\mathrm{K}_{1-p_{\tau, j}}(v_j)}{\mathrm{K}_{-p_{\tau, j}}(v_j)},
\end{split}
\label{eq:variational_moments}
\end{equation}
where $K_p(\cdot)$ is the modified Bessel function of second kind.
}
\IfLongVersion{The calculation of the ELBO $\mathcal{L}$ is provided in Appendix~\ref{sec:appendix_elbo}. }\IfShortVersion{where the parameters of the variational distributions are coupled by their moments. The parameters and moments are provided in Appendix~\ref{sec:appendix}. }Fig~\ref{fig:pgm_mean_field} shows the probabilistic graphical model of the mean field approximation for the generative model. Note the factorization of the random variables which enables tractable posterior inference in comparison to the probabilistic graphical model for the coupled Bayesian regression in Fig.~\ref{fig:pgm}.

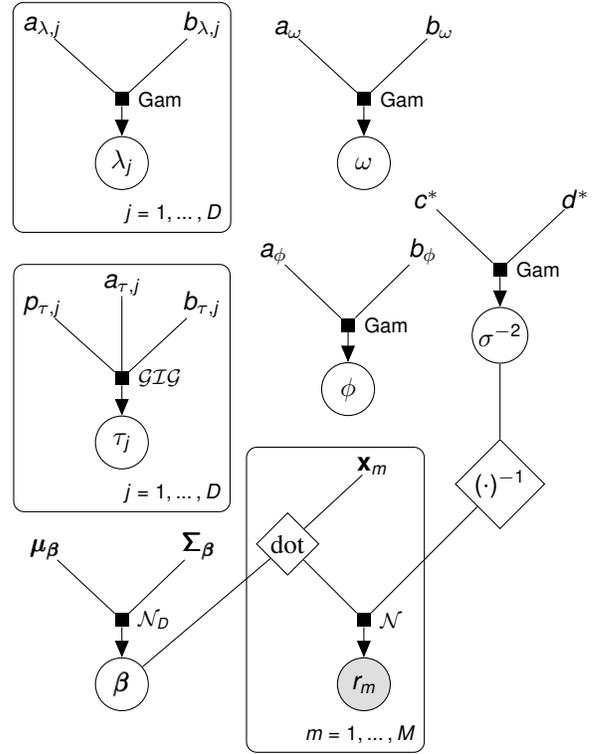
\begin{figure}
  \begin{center}
      \begin{tikzpicture}

  
  \node[obs]	(r){$r_m$};
  \factor[above=of r] {r_f} {right:$\NormalDis$} {} {} ; %
  \node[det, above left= of r_f] (dot_b_x) {dot};
  \node[det, above right=2cm of r_f] (inv) {$(\cdot)^{-1}$};
  \node[const, above right= of dot_b_x] (x_m){$\mathbf{x}_m$};
  \node[latent, left=2.5cm of r ] (beta_j){$\pmb{\beta}$};
  \factor[above = of beta_j]{beta_f} {right:$\mathcal{N}_D$} {} {} ; %
  \node[const, above left= of beta_f](mu_beta) {$\pmb{\mu}_{\pmb{\beta}}$};
   \node[const, above right= of beta_f](Sigma_beta) {$\pmb{\Sigma}_{\pmb{\beta}}$};
   
  \node[latent, above =2.5cm of beta_j](tau_j){$\tau_j$};
  \node[latent,  above =  of inv](sigma){$\sigma^{-2}$};
  \factor[above = of sigma]{sigma_f}{right:$\GammaDis$}{}{};
  \node[const, above left=of sigma_f](c){$c^\ast$};
  \node[const, above right= of sigma_f](d){$d^\ast$};
  
  \factor[above = of tau_j]{tau_f}{right:$\GIGDis$}{}{};
  \node[const, above left = of tau_f] (p_tau) {$p_{\tau,j}$};
  \node[const, above  = of tau_f] (a_tau) {$a_{\tau,j}$};
  \node[const, above right = of tau_f] (b_tau) {$b_{\tau,j}$};
  
  \node[latent, above  =3cm  of tau_j](lambda_j){$\lambda_j$};
  \factor[above = of lambda_j]{lambda_f}{right:$\GammaDis$}{}{};
  \node[const, above left = of lambda_f](a_lambda){$a_{\lambda,j}$};
  \node[const, above right = of lambda_f](b_lambda){$b_{\lambda,j}$}
  ;
  \node[latent,  below right =2.5cm and 2.5cm of lambda_j](phi){$\phi$};
   \factor[above = of phi]{phi_f}{right:$\GammaDis$}{}{};
   \node[const, above left =  of phi_f](a_phi){$a_{\phi}$};
    \node[const, above right =  of phi_f](b_phi){$b_{\phi}$};
    
   \node[latent,  right = 2.5cm of lambda_j](omega){$\omega$};
    \factor[above = of omega]{omega_f}{right:$\GammaDis$}{}{};
      \node[const, above left = of omega_f](halfa){$a_{\omega}$};
        \node[const, above right = of omega_f](halfb){$b_{\omega}$};
        
 \factoredge {dot_b_x,inv} {r_f} {r} ; %
  \edge[-] {beta_j,x_m} {dot_b_x} ;
   \edge[-] {sigma} {inv} ;
    \factoredge {mu_beta,Sigma_beta} {beta_f} {beta_j} ; %
      \factoredge {p_tau,a_tau,b_tau} {tau_f} {tau_j} ; %
       \factoredge {c,d} {sigma_f} {sigma} ; %
        \factoredge {a_lambda,b_lambda} {lambda_f} {lambda_j} ; %
         \factoredge {a_phi,b_phi} {phi_f} {phi} ; %
            \factoredge {halfa,halfb} {omega_f} {omega} ; %
            
   \plate {rx} { %
    (r)(r_f)(r_f-caption) %
    (x_m) %
    (dot_b_x) %
  } {$m=1, \dots, M$} ;
     \plate {} { %
    (tau_j)(tau_f)(tau_f-caption)(p_tau)(a_tau) (b_tau) %
  } {$j= 1, \dots, D$} ;
       \plate {} { %
    (lambda_j)(lambda_f)(lambda_f-caption)(a_lambda)(b_lambda) %
  } {$j= 1, \dots, D$} ;

\end{tikzpicture}
  \end{center}
  \caption{Probabilistic graphical model using a mean field approximation for the Bayesian regression (see Sect.~\ref{subsec:variational_bayes_inference}).
  }
   \label{fig:pgm_mean_field}
\end{figure}

A local optimum of the ELBO $\mathcal{L}$ can be found by cycling through the coupled moments of the variational distributions. This corresponds to a coordinate ascent algorithm on $\mathcal{L}$. \IfLongVersion{The corresponding algorithm is shown in Fig.~\ref{fig:algorithms} Alg.~3.}

\subsection{Stochastic Variational Inference (SVI)}
Next, we present a new posterior inference scheme with TPBN prior based on stochastic variational inference (SVI)~\cite{hoffman2013}. We optimize the ELBO $\mathcal{L}$ by the use of stochastic approximation \cite{robbins1951} where we 
calculate the natural gradient
that is obtained with respect to the natural parameters $\pmb{\eta}$ of the exponential family distribution 
of the mean field variational distributions.

Consider the mean field approximation $q(\pmb{\theta})=\prod_m q(\theta_m)$ for the intractable posterior distribution $p(\pmb{\theta} \mid \mathcal{D})$, where $\pmb{\theta}$ and $\mathcal{D}$ denote the tuple of parameters and the data, respectively. For each factor $q(\theta_m)$, assuming it belongs to the exponential family, the probability density is
$$q(\theta_m)=h(\theta_m) \exp\{\pmb{\eta}^\top \mathbf{S}(\theta_m)-A(\pmb{\eta})\}.$$
Here, $h(\theta_m)$ denotes the base measure, $\pmb{\eta}$ are the natural parameters, $\mathbf{S}(\theta_m)$ is the sufficient statistics of the natural parameters, and $A(\pmb{\eta})$ is the log-normalizer.

We compute the natural gradient of the ELBO $\mathcal{L}$ with respect to the natural parameters of the factorized variational distributions for each variational factor $q(\theta_m)$.
Therefore, the natural gradient computes to \IfLongVersion{
\begin{equation}
\hat{\nabla}_{\pmb{\eta}} \mathcal{L}=\tilde{\pmb{\eta}}-\pmb{\eta},
\label{eq:natural_gradient}
\end{equation}}\IfShortVersion{$\hat{\nabla}_{\pmb{\eta}} \mathcal{L}=\tilde{\pmb{\eta}}-\pmb{\eta}$, }
where $\tilde{\pmb{\eta}}=\mathrm{E}_q[\pmb{\eta}^{\prime}]$.  The parameter $\pmb{\eta}^{\prime}$ is the natural parameter of the full conditional distribution $p(\theta_m \mid \theta_{-m}, \mathcal{D})$, where $\theta_{-m}$ denotes the tuple of all variables but $\theta_m$.  Using a gradient update the variational approximation can be found as \IfLongVersion{
\begin{equation}
\pmb{\eta}^{(n+1)}=\pmb{\eta}^{(n)}+ \gamma_n  \hat{\nabla}_{\eta} \mathcal{L},
\label{eq:gradient_update}
\end{equation}}\IfShortVersion{$\pmb{\eta}^{(n+1)}=\pmb{\eta}^{(n)}+ \gamma_n  \hat{\nabla}_{\eta} \mathcal{L}$, }
where $n$ denotes the iteration step of the algorithm and $\gamma_n$ is a step size parameter.

Random subsampling of the data enables constructing a stochastic approximation algorithm. For this, $\tilde{\pmb{\eta}}$ is replaced by an unbiased estimate $\hat{\pmb{\eta}}$, which yields a stochastic gradient ascent algorithm on the ELBO $\mathcal{L}$ in the form \IfLongVersion{
\begin{equation}
\pmb{\eta}^{(n+1)}=(1-\gamma_n)\pmb{\eta}^{(n)}+\gamma_n \hat{\pmb{\eta}}.
\label{eq:stoch_gradient_update}
\end{equation}}\IfShortVersion{$\pmb{\eta}^{(n+1)}=(1-\gamma_n)\pmb{\eta}^{(n)}+\gamma_n \hat{\pmb{\eta}}$. }
For the step size $\gamma_n$ we use $\frac{1}{n}$.
In the case of the regression problem, we sample one data point $(\mathbf{x}_m,r_m)$ from the set of observed data points and replicate it $M$ times to calculate $\hat{\pmb{\eta}}$. \IfLongVersion{The intermediate estimates of the natural parameters are then obtained by
\begin{equation}
\begin{split}
&\hat{\pmb{\eta}}_{\pmb{\beta}}=\left(\langle\sigma^{-2}\rangle M \mathbf{x}_m r_m, -\frac{1}{2} \langle \sigma^{-2}\rangle (M \mathbf{x}_m \mathbf{x}_m^\top+\mathbf{T}^{-1})\right),\;  \\
&\hat{\pmb{\eta}}_{\sigma^{-2}}=\left(\frac{M+D+c_0}{2}-1, -(M r_m^2 - 2M r_m \mathbf{x}_m^\top \langle \pmb{\beta}\rangle\right.\\
&\qquad\qquad + \left.M \mathbf{x}_m^\top \langle \pmb{\beta}\pmb{\beta}^{\top}\rangle \mathbf{x}_m  +\sum_{j=1}^D \langle \beta_j^2 \rangle \langle \tau_j^{-1} \rangle+d_0)/2\right),\;  \\
&\hat{\pmb{\eta}}_{\tau, j}=\left(a_0-\frac{3}{2}, -\langle \lambda_j \rangle, \langle \beta_j^2 \rangle \langle\sigma^{-2}\rangle/2\right),\;\\
&\hat{\pmb{\eta}}_{\lambda,j}=\left(a_0+b_0-1, -\langle\tau_j\rangle-\langle\phi\rangle\right),\;\\
& \hat{\pmb{\eta}}_{\phi}=\left(D b_0 -\frac{1}{2},-\langle\omega\rangle -\sum_{j=1}^D \langle\lambda_j\rangle\right),\;\\
&\hat{\pmb{\eta}}_{\omega}=\left(0,-\langle\phi\rangle-1\right).
\end{split}
\label{eq:intermediate_estimates}
\end{equation}
The derivation is provided in Appendix \ref{sec:appendix_intermediate}.

The transformation from the natural parametrization to the variational parametrization is calculated using
\begin{equation}
\begin{split}
&\left(\pmb{\mu}_{\pmb{\beta}}, \pmb{\Sigma}_{\pmb{\beta}}\right)=\left(-\frac{1}{2}{{\pmb{\eta}^{(2)}_{\pmb{\beta}}}}^{-1}{\pmb{\eta}^{(1)}_{\pmb{\beta}}},-\frac{1}{2} {{\pmb{\eta}^{(2)}_{\pmb{\beta}}}}^{-1}\right),\; \\
&\left(c^\ast,d^\ast\right)=\left( {{{\eta}^{(1)}_{\sigma^{-2}}}}+1,- {{{\eta}^{(2)}_{\sigma^{-2}}}}\right),\;\\
&\left(p_{\tau, j}, a_{\tau, j}, b_{\tau, j}\right)=\left({{\eta}^{(1)}_{\tau, j}}+1,-2{{\eta}^{(2)}_{\tau, j}}, 2{{\eta}^{(3)}_{\tau, j}}\right),\;\\
&\left(a_{\lambda,j}, b_{\lambda,j}\right)=\left({\eta^{(1)}_{\lambda,j}}+1, -{\eta^{(2)}_{\lambda,j}}\right),\;\\
& \left(a_{\phi}, b_{\phi}\right)=\left({\eta^{(1)}_{\phi}}+1, -{\eta^{(2)}_{\phi}}\right),\;\\
&\left(a_{\omega}, b_{\omega}\right)=\left({\eta^{(1)}_{\omega}}+1, -{\eta^{(2)}_{\omega}}\right)
\end{split}
\label{eq:natural_transform}
\end{equation}
and the moments can then be calculated with \eqref{eq:variational_moments}. We denote by $\eta^{(i)}$ the $i$-th variable of the tuple of natural parameters $\pmb{\eta}$.}
The gradient update\IfLongVersion{ \eqref{eq:stoch_gradient_update}} with random subsampling is performed until the ELBO $\mathcal{L}$ converges. \IfLongVersion{For an algorithmic description of SVI see Fig.~\ref{fig:algorithms} Alg.~2.}\IfShortVersion{The intermediate estimates $\hat{\pmb{\eta}}$ are provided in Appendix~\ref{sec:appendix}.}

\begin{figure}
	\centering
	\includegraphics[width=0.9\columnwidth]{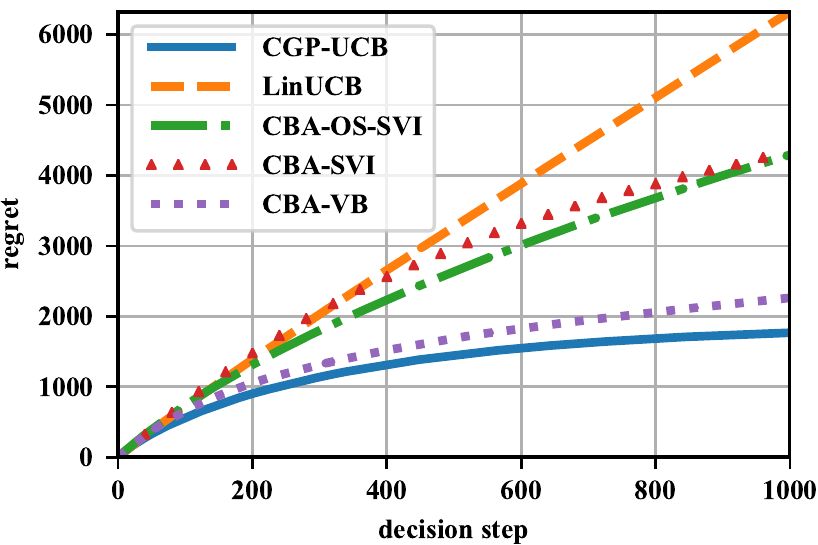}
	\caption{Average regret for our contextual bandit algorithms vs. the baseline (CGP and LinUCB) for a dense linear model.}
	\label{fig:regret_full}
\end{figure}

\subsection{One Step Stochastic Variational Inference (OS-SVI)}
Since the optimization of the ELBO $\mathcal{L}$ until convergence with both VB and SVI is computationally expensive, we present a novel one-step SVI (OS-SVI) algorithm for the bandit setting.
In each round of OS-SVI the learner observes a context and a reward $(\mathbf{x}_{t}(a), r_{t}(a))$ based on the taken  action $a_t=a$. This data point is used to update the variational parameters of the $a$-th regression coefficients $\pmb{\beta}(a)$ by going one step in the direction of the natural gradient of the ELBO $\mathcal{L}$. For this we calculate the intermediate estimates \IfLongVersion{\eqref{eq:intermediate_estimates}} based on $t$ replicates of the observed data point $(\mathbf{x}_{t}(a),r_{t}(a))$. Thereafter, the stochastic gradient update is performed\IfLongVersion{ with \eqref{eq:stoch_gradient_update}}. By transforming the natural parameters back to their corresponding parametric form \IfLongVersion{\eqref{eq:natural_transform}}, the updated mean $\pmb{\mu}_{\pmb{\beta}(a)}$ and covariance matrix $\pmb{\Sigma}_{\pmb{\beta}(a)}$ can be found. This update step is computationally significantly faster than using VB or SVI. \IfLongVersion{The  OS-SVI subroutine is described in Fig.~\ref{fig:algorithms} Alg.~4.}

\subsection{Accuracy and Computational Speed of the CBA-UCB Algorithms}
For the numerical evaluation of the CBA-UCB with three parameter Beta Normal prior, we first create data based on the linearization assumption. We use a problem with decision horizon $T=1000$, $D=20$ dimensions, and $K=20$ actions. We use two experimental setups with a dense regression coefficient vector and a sparse regression coefficient vector, i.e., only five regression coefficients are unequal to zero.

\begin{figure}
\centering
\includegraphics[width=0.9\columnwidth]{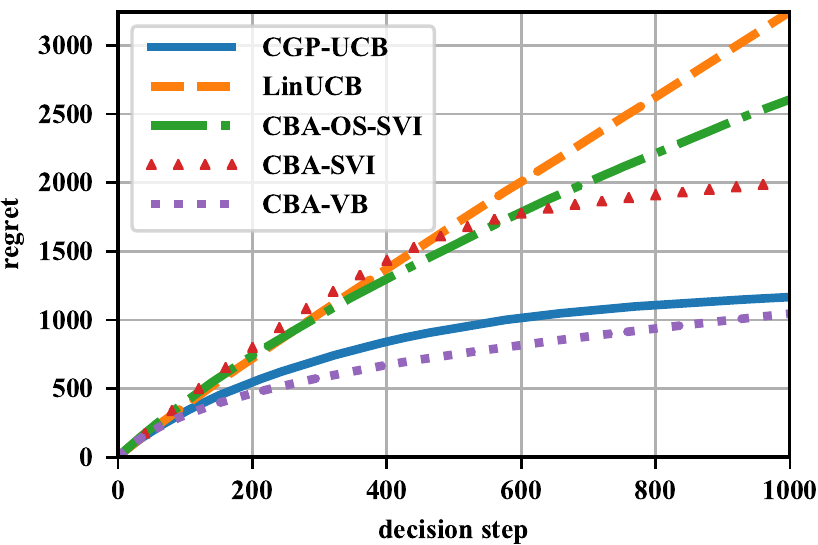}
\caption{Average regret for our contextual bandit algorithms vs. the baseline (CGP and LinUCB) for a sparse linear model.}
\label{fig:regret_sparse}
\end{figure}

We compare the developed algorithm CBA-UCB using the variants VB, SVI and OS-SVI with two base-line algorithms: LinUCB \cite{li2010} and CGP-UCB \cite{krause2011}. For the CGP-UCB, we use independent linear kernels for every action. Fig.~\ref{fig:regret_full} and Fig.~\ref{fig:regret_sparse} show the average regret \eqref{eq:regret} for the dense and the sparse setting, respectively. For the sparse setting expected in high-dimensional problems such as network-assisted video streaming, CBA-UCB with VB yields the smallest regret. We observe in Fig.~\ref{fig:regret_full} that in the dense setting CGP-UCB obtains a high performance which is closely followed by CBA-UCB with VB. Note that CGP-UCB performs well, as Gaussian process regression with a linear kernel corresponds to a dense Bayesian regression with marginalized regression coefficients, and therefore matches the model under which the dense data has been created.

In Fig~\ref{fig:table} we show the run-times of the algorithms, where we observe that the run-times for CBA-UCB with VB / SVI and the CGP-UCB baseline are impractically high. Further, this running performance deteriorates as the dimensions of the context grow, since the computational bottleneck of both VB and SVI are multiple matrix inversions of size $D \times D$\IfLongVersion{, see Fig.~\ref{fig:comp_time_D}.  Fig.~\ref{fig:comp_time_T} shows the scaling of the run-time with the decision horizon $T$ with an identical setup as in Tab.~\ref{fig:table}. The CGP-UCB scales badly with $T$, as the kernel matrix of size $M_a \times M_a$ is inverted at every time step. Here, $M_a$ denotes the number of already observed contexts and rewards for decision $a$}. Since the decision making has to be made in the order of a few hundred milliseconds for video streaming applications, neither CBA-UCB with VB nor CGP-UCB can be computed within this timing restriction. Therefore, we resort to the OS-SVI variant of the CBA algorithm, which empirically obtains a much smaller regret than the fast LinUCB baseline algorithm, but still retains a comparable run-time performance\footnote{For updating CBA-UCB with OS-SVI  or LinUCB we only have to invert a $D \times D$ matrix once after a decision.}. This renders the use of CBA with One Step Stochastic Variational Inference for network-assisted video quality adaptation feasible.

\renewcommand{\figurename}{Tab.}
\setcounter{figure}{0}
\begin{figure}
\centering
\begin{tabular}{@{}llr@{}} \toprule
Algorithm & Sparse Setting & Dense Setting\\ \midrule
CGP-UCB  & 638.68 s &  643.44 s\\
LinUCB   & 31.24 s & 30.70 s  \\
CBA-OS-SVI   & 91.40 s   & 89.56  s \\
CBA-SVI & 3784.00 s & 4081.74 s \\
CBA-VB &  1434.11 s & 1760.83 s \\
 \bottomrule
\end{tabular}

\caption{Run-times for $N=100$ simulations of the CBA algorithms compared to the baseline algorithms CGP-UCB and LinUCB. Simulations executed on an Intel\textsuperscript{\textregistered} Xeon\textsuperscript{\textregistered} E5-2680 v3 @2.5GHz machine.}
\label{fig:table}
\end{figure}
\renewcommand{\figurename}{Fig.}
\setcounter{figure}{6}

\IfLongVersion{
\begin{figure}
\centering
\includegraphics[width=\columnwidth]{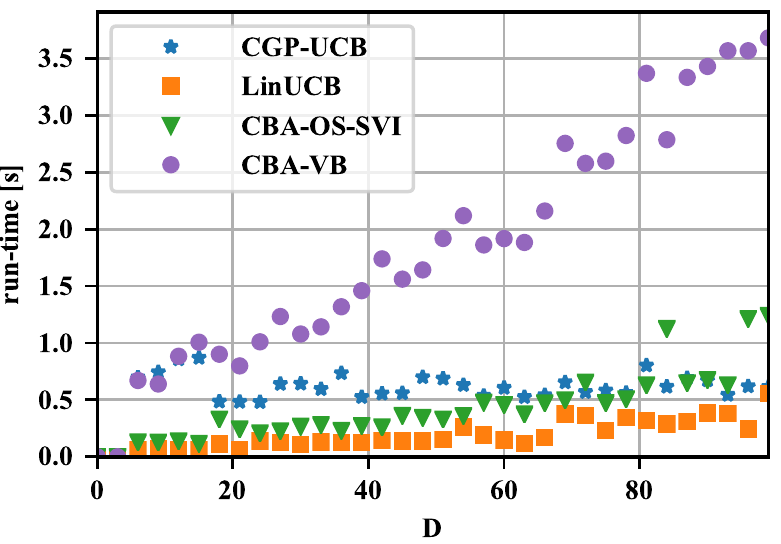}
\caption{Run-time vs. context dimensions $D$ for a sparse linear model, with $K=20$ actions and a decision horizon of $T=100$. CBA-UCB with SVI not shown for clarity. }
\label{fig:comp_time_D}
\end{figure}

\begin{figure}
\centering
\includegraphics[width=\columnwidth]{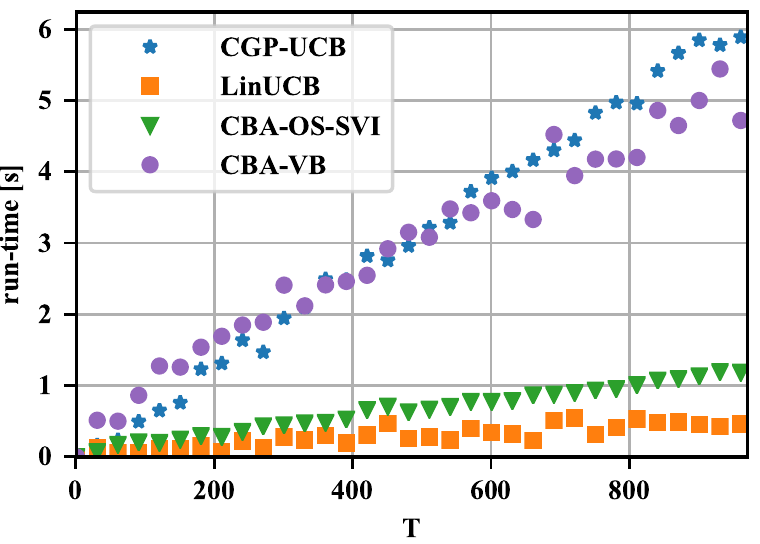}
\caption{Run-time vs. decision horizon $T$ for a sparse linear model, with $K=20$ actions and $D=20$ features. CBA-UCB with SVI not shown for clarity.}
\label{fig:comp_time_T}
\end{figure}
} 
\section{Video Quality Adaptation as a Contextual Bandit Problem}
\label{sec:video_adaptation_as_CB}

In the following, we model ABR streaming as a contextual bandit problem where we use our developed CBA algorithm for video quality adaptation. The action set corresponds to the set of available bitrates $\mathcal{V}$ such that action $a_t \in \{1,\dots,K\}$ represents the decision to request quality $v(a_t)$ for the $t$-th segment; i.e., to request the segment $s_{t}(a_t)$. Below we formulate a real-valued segment-based QoE function to represent the reward $r_{t}(a_t)$ obtained by performing $a_t$. Furthermore, we let $\mathbf{x}_{t}(a)$ represent the network context vector corresponding to an available action $a$ at segment $t$. At each $t$, therefore, there will be $K$ unique context vectors available.

\subsection{Online Video Quality Adaptation using CBA}

CBA performs online video quality adaptation by calculating the index presented in \eqref{eq:idx_gaussian} for each available action after observing the context vector $\mathbf{x}_{t}(a)$ of the action to determine the optimal bitrate to request for the next segment $t$.
There are no constraints on the contents of the context vectors, allowing CBA to learn with any information available in the networking environment.
Furthermore, each context feature may be either global or action-specific; for example, the current buffer filling percentage or the last 50 packet RTTs at bitrate $v(a)$, respectively. The action $a_t$ with the largest computed index is chosen, and a request goes out for $s_{t}(a_t)$. Once $s_{t}(a_t)$ is received, its QoE value below is calculated and fed to CBA as the reward $r_{t}(a_t)$. CBA then updates its internal parameters before observing the next set of context vectors and repeating the process for segment $t+1$, until the video ends at segment $T$.

The performance of CBA depends upon several hyperparameters. In the description in \IfShortVersion{Sect.~\ref{sec:cba_ucb}}\IfLongVersion{Fig.~\ref{fig:algorithms}, Alg.~1.}, we choose $\alpha=1$ as it was shown to yield the most promising results \cite{kaufmann2012}. As mentioned in Sect~\ref{sec:multi_armed_bandits}, we use $a_0=b_0=1/2$ to obtain the horseshoe shrinkage prior. We let $c_0=d_0=10^{-6}$; we choose $c_0$ and $d_0$ to be small nonzero values such that a vague prior is obtained.

\subsection{Reward Formulation: Objective QoE}
The calculated QoE metric is the \emph{feedback} used by CBA to optimize the quality adaptation strategy.
As QoE scores for a video segment may vary among users, we resort in this work to an objective QoE metric similar to~\cite{Yin:2015} which is
derived from the following set of factors:

\begin{enumerate}
    \item \textbf{Video quality}: The bitrate of the segment. $v(a_t) \in \mathcal{V}$.
    \item \textbf{Decline in quality}: If the current segment is at a lower bitrate than the previous one, $[v(a_{t-1}) - v(a_t)]_+$ for two back to back segments\footnote{we use $[x]_+$ to denote $\max\{x,0\}$.}.
    \item \textbf{Rebuffer time}: The amount of time spent with an empty buffer after choosing $v(a_t)$.
\end{enumerate}

The rationale behind using the decline in quality, in contrast to the related work that counts quality variations, is that we do not want to penalize CBA if the player strives for higher qualities without risk of rebuffering.
The importance of each component may vary based on the specific user or context, so, similar to \cite{Yin:2015}, we define the QoE of a segment as a weighted sum of the above factors. Let the rebuffer time $G(v(a_t))$ be the amount of time spent rebuffering after choosing $v(a_t)$. 
We define the QoE then as:
\begin{equation}
\begin{split}
\text{QoE}&(s_{t}(a_t)) = w_1 v(a_t) - w_2[v(a_{t-1}) - v(a_t)]_+  - w_3G(v(a_t))
\end{split}
\label{eq:qoe}
\end{equation}
where $w_1$, $w_2$, and $w_3$ are non-negative weights corresponding to the importance of the video quality, decline in quality, and rebuffer time, respectively. For a comparison of several instantiations of these weights, see \cite{Yin:2015}.

Note that the above QoE metric is independent from CBA; the bandit is only given the scalar result of the calculation. \emph{CBA is able take arbitrary QoE metrics as specified input as long as these comprise a real-valued function to produce the reward metric.}

\section{Evaluation of Quality Adaptation in NDN}\label{sec:evaluation}

\begin{figure}
	\includegraphics[width=0.9\columnwidth]{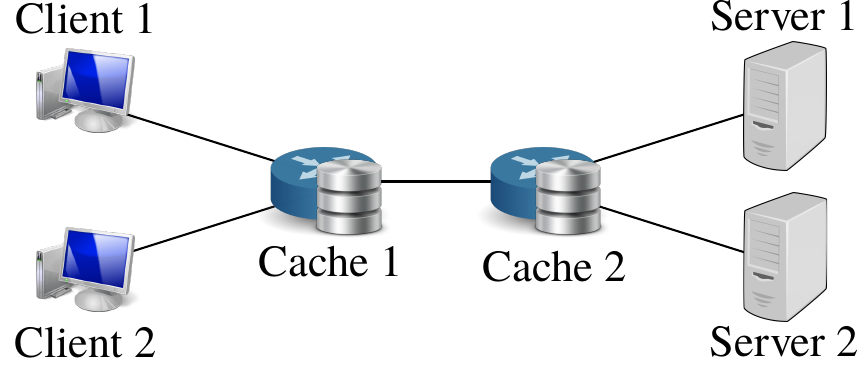}
	\caption{Emulation testbed for the \textit{doubles} topology. Client link capacity follows a bandwidth trace, server links have a capacity of 20Mbps, and the internal cache link has a capacity of 1000Mbps. Caches can store up to 1500 Data chunks.}
	\label{fig:topo}
\end{figure}

\begin{figure}
	\includegraphics[width=0.9\columnwidth]{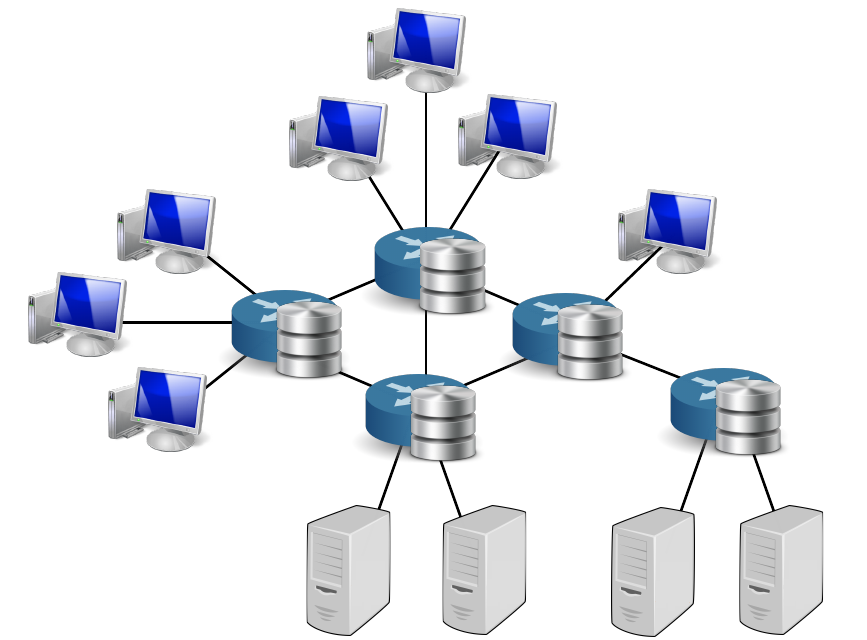}
	\caption{Emulation testbed for the \textit{full} topology. Client and server links have a capacity of 20Mbps, and the internal cache links have a capacity of 1000Mbps. Caches can store up to 1500 Data chunks.}
	\label{fig:topo_full}
\end{figure}

To evaluate the performance of CBA and compare it with Throughput-based (TBA) and Buffer-based (BBA) adaptation peers, we emulate the two NDN topologies: the \textit{doubles} topology, shown in Fig.~\ref{fig:topo}; and the \textit{full} topology, shown in Fig.~\ref{fig:topo_full}. The topologies are built using an extension of the Containernet project\footnote{https://github.com/containernet/containernet} which allows the execution of Docker hosts as nodes in the Mininet emulator.

The NDN clients use a DASH player implemented with \texttt{libdash}, based on the code from \cite{samain2017}
with Interest Control Protocol (ICP) parameters of $\gamma_{\text{ICP}}=2$, $\beta_{\text{ICP}}=0.5$, and $\texttt{initialWindow}=300$. We note that traffic burstiness can vary significantly depending on the ICP parameters used.

The clients begin playback simultaneously, where they stream the first 200 seconds of the BigBuckBunny video encoded in two-second H.264-AVC segments offered at the $K=5$ quality bitrates \{1, 1.5, 2.1, 3, 3.5\}Mbps, with a playback buffer size of 30 seconds. All containers run instances of the NDN Forwarding Daemon (NFD) with the \texttt{access} strategy, and repo-ng is used to host the video on the servers and caches.

\begin{figure}
	\centering
	\includegraphics[width=0.9\columnwidth]{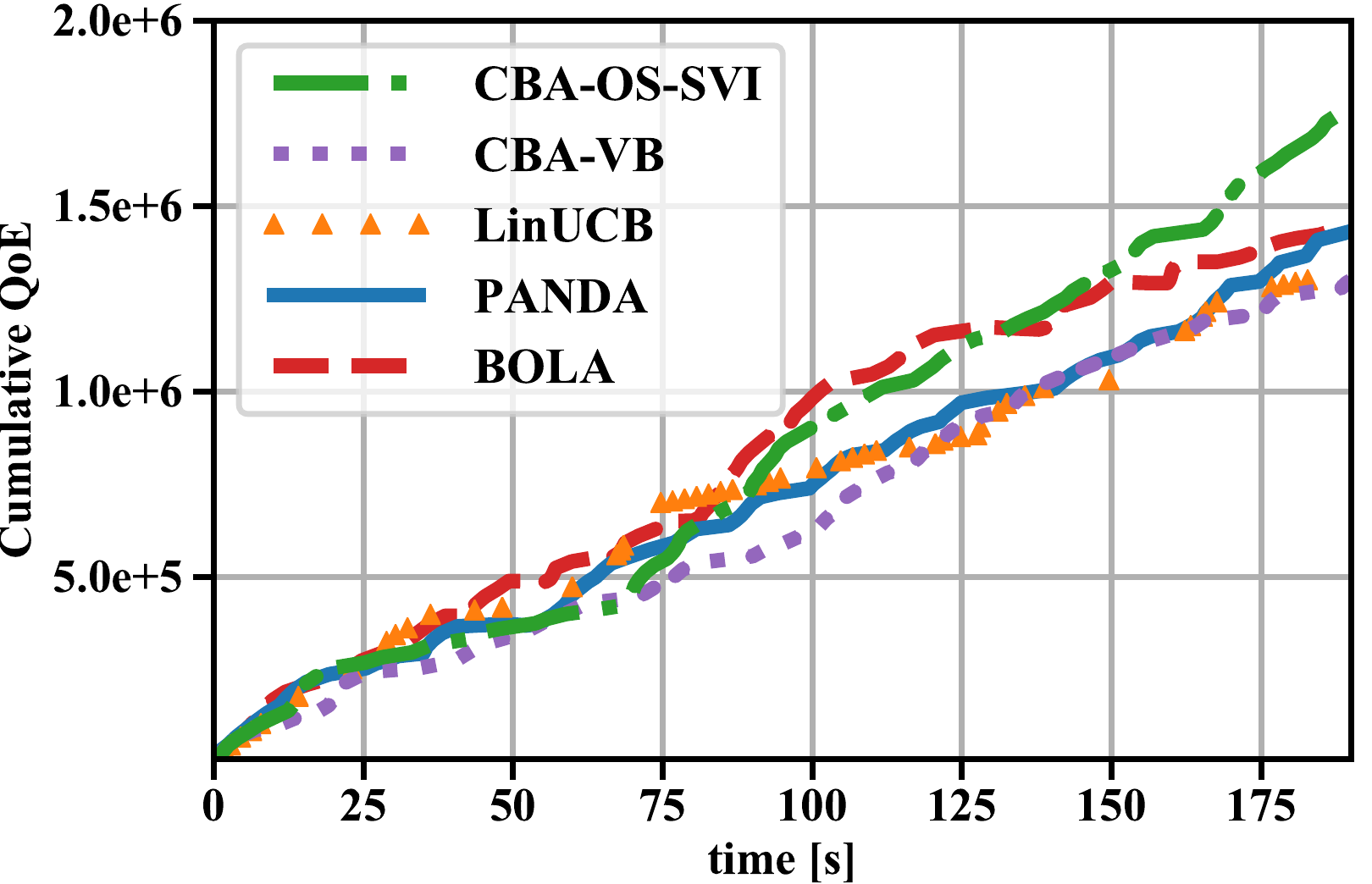}
	\caption{Client 1 QoE over playback on the doubles topology.}
	\label{fig:cumul_qoe}
\end{figure}

\begin{figure}
	\centering
	\includegraphics[width=0.9\columnwidth]{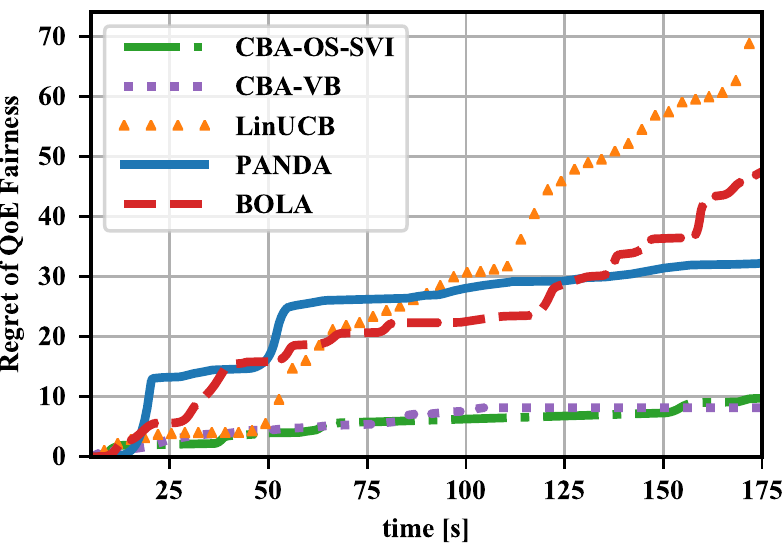}
	\caption{QoE fairness evaluation on the doubles topology.}
	\label{fig:fairness}
\end{figure}

In the following, we compare the performance of CBA in the VB and OS-SVI variants, in addition to the baseline algorithm LinUCB \cite{li2010}. We also examine the performance of two state-of-the-art BBA and TBA algorithms, i.e., BOLA~\cite{Bola16} and PANDA~\cite{Panda14}, respectively. There are many adaptation algorithms in the literature, some of which use BBA and TBA simultaneously, including \cite{Yin:2015}, \cite{mao2017}, \cite{akhtar2018}, and \cite{bentaleb2018}; however, BOLA and PANDA were chosen because they are widely used and achieve state-of-the-art performance in standard HTTP environments. Buffer filling percentage and quality-specific segment packet RTTs are provided to the client as context. Furthermore, we added a \texttt{numHops} tag to each Data packet to track the number of hops from the Data origin to the consumer.

We track the RTTs and number of hops of the last 50 packets of each segment received by the client in accordance with measurements from \cite{Wang:2017}. If a segment does not contain 50 packets, results from existing packets are resampled. As a result, each CBA algorithm is given a $D=101$ dimensional context vector constituted of the buffer fill percentage, packet RTTs, and \texttt{numHops} for each of the $K=5$ available qualities.

\renewcommand{\figurename}{Tab.}
\setcounter{figure}{0}
\begin{figure}
	\centering
	\begin{tabular}{@{}llllr@{}} \toprule
	Algorithm & \thead{\shortstack{Bitrate\\ {[}Mbps{]}}} & \thead{\shortstack{Quality\\ switches\\ {[}\#{]}}} & \thead{\shortstack{Switch\\ magnitude\\ {[}Mbps{]}}} & \thead{\shortstack{Parameter\\ update\\ time {[}ms{]}}}\\ \midrule
	CBA-OS-SVI & \textbf{3.10} & \textbf{6} & \textbf{0.57} & 15 \\
	CBA-VB & 2.58 & \textbf{6} & 0.65 & 325 \\
	LinUCB   & 2.24 & 14 & 1.07 & \textbf{6} \\
	BOLA & 2.63 & 36 & 1.19 & \\
	PANDA & 2.51 & 16 & 1.00 & \\
	\bottomrule
\end{tabular}
	\caption{Client 1 streaming statistics on the doubles topology.
	}
	\label{fig:table_eval}
\end{figure}
\renewcommand{\figurename}{Fig.}
\setcounter{figure}{10}

\subsection{Results on the Doubles Topology}

We modulate the capacity of the bottleneck link using truncated normal distributions. The link capacity is hence drawn with mean of 7Mbps, where it stays unchanged for a period length drawn with a mean of 5s. The weights in Eq.~\ref{eq:qoe} are set to $w_1=6$, $w_2=2$, and $w_3=2$, emphasizing the importance of the average quality bitrate without allowing a large amount of rebuffering to take place. We note that the use of subjective quality evaluation tests for different users to map these weights to QoE metrics via, e.g., the mean opinion score (MOS), is out of the scope of this work. 

Examining Tab.~\ref{fig:table_eval}, we see that the one-step CBA-OS-SVI yields a significantly higher average bitrate. This is expected based on the QoE definition \eqref{eq:qoe}, but we might expect CBA-VB to pick high bitrates as well. However, we observe that the parameter update time for CBA-VB is 20 times greater than that of CBA-OS-SVI; this puts a delay of one-sixth of each segment length on average between receiving one segment and requesting another.
Looking at CBA-VB in Fig.~\ref{fig:cumul_qoe} we see that CBA-VB accumulates a much larger rebuffer time than other methods. Hence, CBA-VB is forced to request lower bitrates to cope with the extra rebuffer time incurred by updating its parameters.
%
In addition, note that LinUCB fails to select high bitrates despite having a very small parameter update time, implying that LinUCB is not adequately fitting the context to the QoE and is instead accumulating a large amount of regret. This is corroborated by its cumulative QoE depicted in Fig.~\ref{fig:cumul_qoe}, which performs nearly as poorly as CBA-VB. By inducing sparsity on the priors and using just one sample, CBA-OS-SVI successfully extracts the most salient features quickly enough to obtain the highest cumulative QoE of all algorithms tested.

Interestingly, the CBA approaches shown in Fig.~\ref{fig:table_eval} also result in the lowest number of quality switches, though our QoE metric does not severely penalize quality variation. We see that the magnitude of their quality switches is also nearly half that of the other algorithms.

%

Concerning the rebuffering behavior, we observe rebuffering ratios of \{4.5\%, 8.4\%, 11.4\%, 17.6\%, 32.9\%\} for LinUCB, BOLA, PANDA, CBA-OS-SVI, and CBA-VB, respectively. We trace some of the rebuffering events to the ICP congestion control in NDN. Note that tuning the impact of rebuffering on the adaptation decision is not a trivial task \cite{samain2017}. Fortunately, this is not hardwired in CBA but rather given through \eqref{eq:qoe}. Hence, in contrast to state-of-the-art adaptation algorithms, CBA could learn to filter the contextual information that is most important for rebuffering by tweaking the QoE metric used.

An important consideration when choosing a quality adaptation algorithm is fairness among clients while simultaneously streaming over common links.
While this is taken care of in DASH by the underlying TCP congestion control, we empirically show here how the ON-OFF segment request behavior, when paired with the considered quality adaptation algorithms, impacts the QoE fairness in NDN.
This is fundamentally different from considering bandwidth sharing fairness in NDN; e.g., in \cite{samain2017}. Here we are interested in QoE fairness since \emph{the QoE metric and not the bandwidth share is the main driver of the quality adaptation algorithm}.
Fig.~\ref{fig:fairness} shows the regret of QoE fairness between both clients
, where a larger regret indicates a greater difference in QoE between both clients up to a particular segment.
Note that the regret is defined as a cumulative metric similar to~\eqref{eq:regret}.
In accordance to the discussion in\cite{lan2010}, the fairness measure used here is the entropy of the relative QoE of the two clients $H_B\left(\frac{\mathrm{QoE}_\mathrm{client 1}(t)}{\mathrm{QoE}_\mathrm{client 1}(t)+\mathrm{QoE}_\mathrm{client 2}(t)}\right),$ where $H_B(\cdot)$ denotes the binary entropy and the QoE is given by \eqref{eq:qoe}.
The regret is calculated with respect to the optimal fairness of  $H_B^{\ast}\left(\frac{\mathrm{QoE}_\mathrm{client 1}(t)}{\mathrm{QoE}_\mathrm{client 1}(t)+\mathrm{QoE}_\mathrm{client 2}(t)}\right)=1$.
Observe that the CBA algorithms attain a significantly lower QoE fairness regret than other techniques.

\subsection{Results on the Full Topology}

To evaluate the capacity of CBA to adapt to different reward functions in complex environments, we compare performance with the full topology on two sets of weights in Eq.~\ref{eq:qoe}: \texttt{HIGH\_QUALITY\_WEIGHTS} sets $w_1=6$, $w_2=2$, and $w_3=2$, identical to those used in the evaluation on the doubles topology; conversely, \texttt{NO\_REBUFFERING\_WEIGHTS} sets $w_1=1$, $w_2=1$, and $w_3=3$, placing greater importance on continuous playback at the expense of video quality. We evaluate each algorithm with each weighting scheme for 30 epochs, where one epoch corresponds to streaming 200 seconds of the BigBuckBunny video. All clients use the same adaptation algorithm and weighting scheme within an epoch, and bandits begin each epoch with no previous context information.

Inspecting Tab.~\ref{fig:table_eval_full}, we observe that the performance statistics among algorithms, even with different weighting schemes, are much closer than for the doubles topology. We attribute this to the use of a more complicated topology in which many more clients are sharing network resources, resulting in fewer and less predictable resources for each client. Furthermore, the average bitrate for the bandit algorithms does not change significantly across weighting schemes, and either stays the same or increases when using \texttt{NO\_REBUFFERING\_WEIGHTS}. This may seem contradictory, but, analyzing part (a) of Figs.~\ref{fig:full_hq} and ~\ref{fig:full_nr}, we note that CBA-OS-SVI tended to choose much lower bitrates with \texttt{NO\_REBUFFERING\_WEIGHTS}, and therefore accruing less rebuffer time in part (b), than with \texttt{HIGH\_QUALITY\_WEIGHTS}, indicating that CBA-OS-SVI successfully adapted to either weighting scheme within the playback window. Similarly to the doubles topology, LinUCB failed to map the context to either weighting scheme, selecting higher bitrates and rebuffering longer with \texttt{NO\_REBUFFERING\_WEIGHTS}. Note that, for either CBA-OS-SVI or LinUCB, the cumulative rebuffer time in part (b) of Figs.~\ref{fig:full_hq} and ~\ref{fig:full_nr} tapers off roughly halfway through the video, as either algorithm learns to request more appropriate bitrates.

Interestingly, CBA-VB also fails to adapt to either weighting scheme, performing nearly identically in either case. This is a byproduct of the excessive parameter update time for CBA-VB in Tab.~\ref{fig:table_eval_full}, which stems from the unpredictable nature of a larger network and the computational strain of performing up to 7 CBA-VB parameter updates simultaneously on the test machine. CBA-VB is therefore spending over half of the length of each segment deciding on which segment to request next, causing long rebuffering times in part (b) of Figs.~\ref{fig:full_hq} and ~\ref{fig:full_nr}, culminating in very low QoE scores regardless of the weighting scheme used. This obfuscates the underlying QoE function, preventing CBA-VB from differentiating between the weights in either case within the time allotted. In a real-world scenario, where each client is an independent machine, we expect that CBA-VB, as well as CBA-OS-SVI and LinUCB to a lesser extent, would have parameter update times comparable to those in the doubles topology, resulting in better performance; however, we note that evaluation in such an environment is out of the scope of this work.

Again, we see in Tab.~\ref{fig:table_eval} that CBA-OS-SVI switches qualities least frequently despite neither weighting scheme explicitly penalizing quality variation. Furthermore, according to parts (c) and (d) of Fig.~\ref{fig:full_hq} and Fig.~\ref{fig:full_nr}, CBA-OS-SVI and CBA-VB are both stable in the number of quality switches and the quality switch magnitude across epochs, even under different weighting schemes, as opposed to the other algorithms tested.

\renewcommand{\figurename}{Tab.}
\setcounter{figure}{1}
\begin{figure}
	\centering
	\begin{tabular}{@{}llllr@{}} \toprule
	Algorithm & \thead{\shortstack{Bitrate\\ {[}Mbps{]}}} & \thead{\shortstack{Quality\\ switches\\ {[}\#{]}}} & \thead{\shortstack{Switch\\ magnitude\\ {[}Mbps{]}}} & 
	\thead{\shortstack{Parameter\\ update\\ time {[}ms{]}}}\\ 
	\midrule
	\multicolumn{5}{@{}l}{\texttt{HIGH\_QUALITY\_WEIGHTS}} \\
	\midrule
	CBA-OS-SVI & 1.55 & \textbf{5} & 0.82 & 53 \\
	CBA-VB & 1.52 & 15 & 1.16 & 1254 \\
	LinUCB & 1.27 & 17 & 1.01 & \textbf{11} \\
	BOLA & \textbf{1.96} & 8 & 0.63 & \\
	PANDA & 1.15 & 18 & \textbf{0.56} & \\
	\midrule
	\multicolumn{5}{@{}l}{\texttt{NO\_REBUFFERING\_WEIGHTS}} \\
	\midrule
	CBA-OS-SVI & 1.55 & \textbf{6} & 0.93 & 55 \\
	CBA-VB & 1.68 & 12 & 1.08 & 1362 \\
	LinUCB & 1.43 & 22 & 1.04 & \textbf{16} \\
	BOLA & \textbf{1.92} & 12 & 0.71 & \\
	PANDA & 1.13 & 17 & \textbf{0.70} & \\
	\bottomrule
\end{tabular}
	\caption{Client 1 streaming statistics on the full topology.
	}
	\label{fig:table_eval_full}
\end{figure}
\renewcommand{\figurename}{Fig.}
\setcounter{figure}{12}

\IfLongVersion{
	%
	
	\begin{figure*}
		\centering
		\begin{subfigure}{0.48\textwidth}
			\includegraphics[width=\linewidth, height=4.6cm]{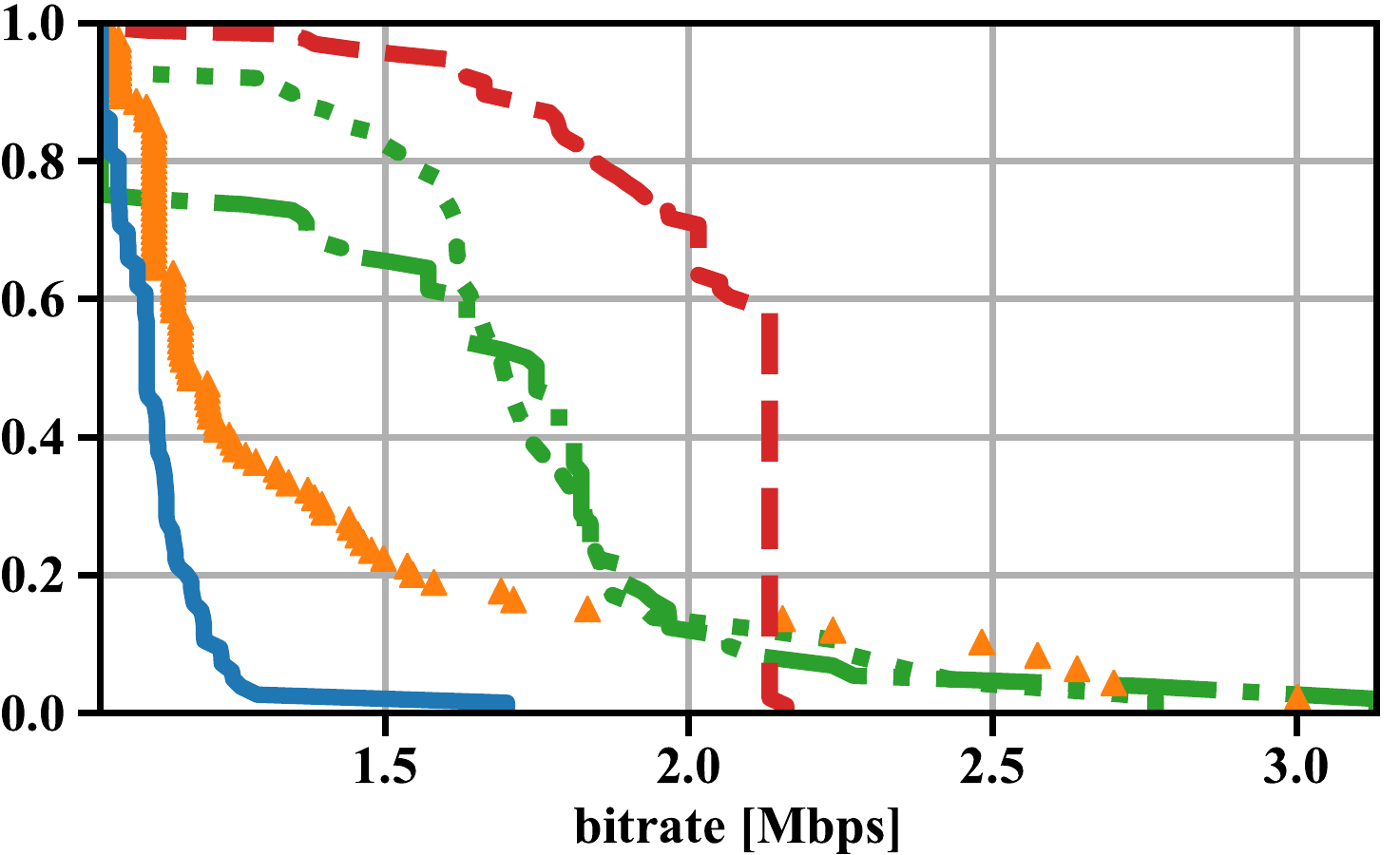}
			\caption{CCDF of average bitrate chosen per epoch.}
		\end{subfigure}\hspace*{\fill}
		\begin{subfigure}{0.48\textwidth}
			\includegraphics[width=\linewidth, height=4.6cm]{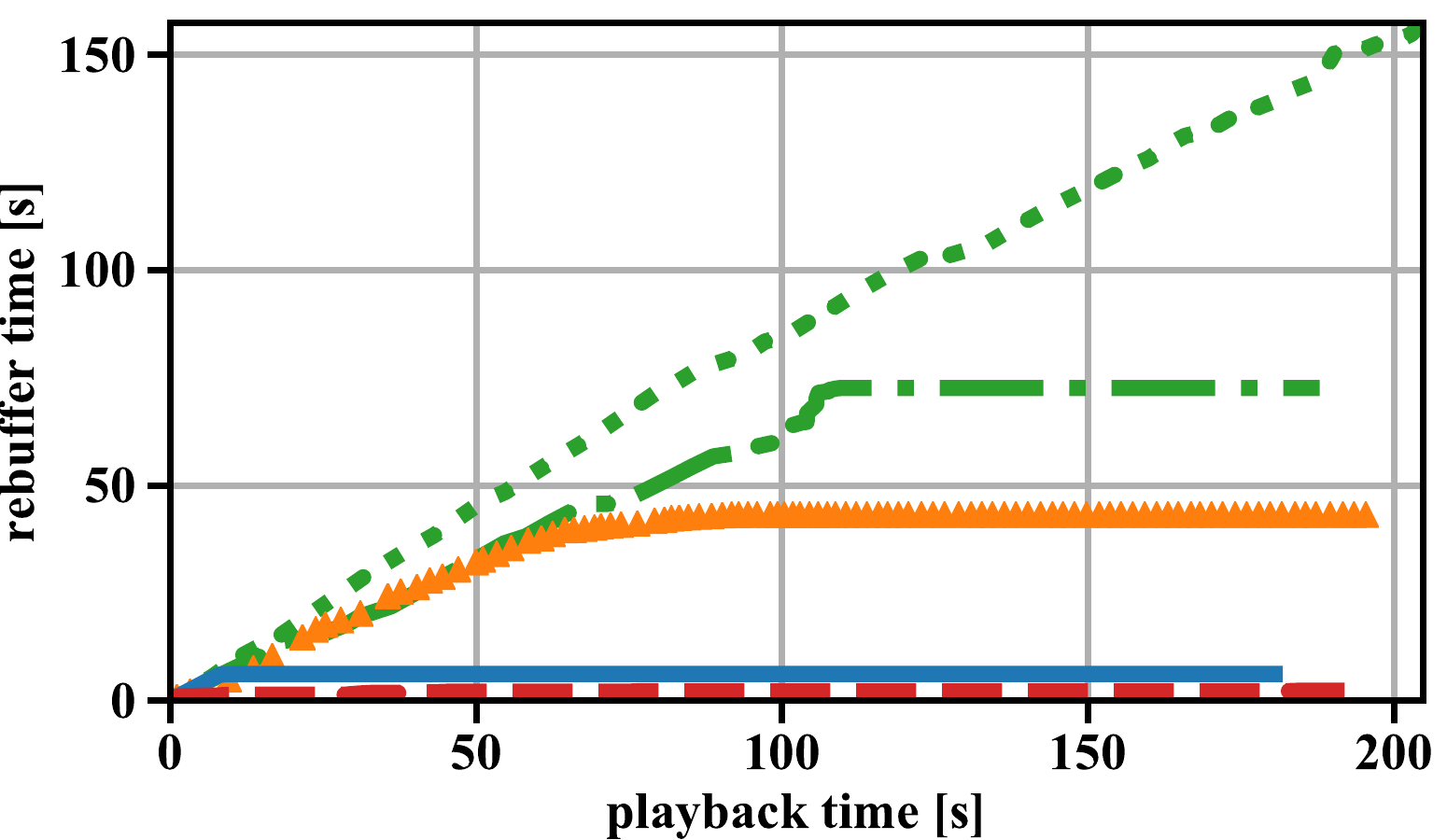}
			\caption{Average cumulative rebuffer time during playback.}
		\end{subfigure}
		
		\medskip
		\begin{subfigure}{0.48\textwidth}
			\includegraphics[width=\linewidth, height=4.6cm]{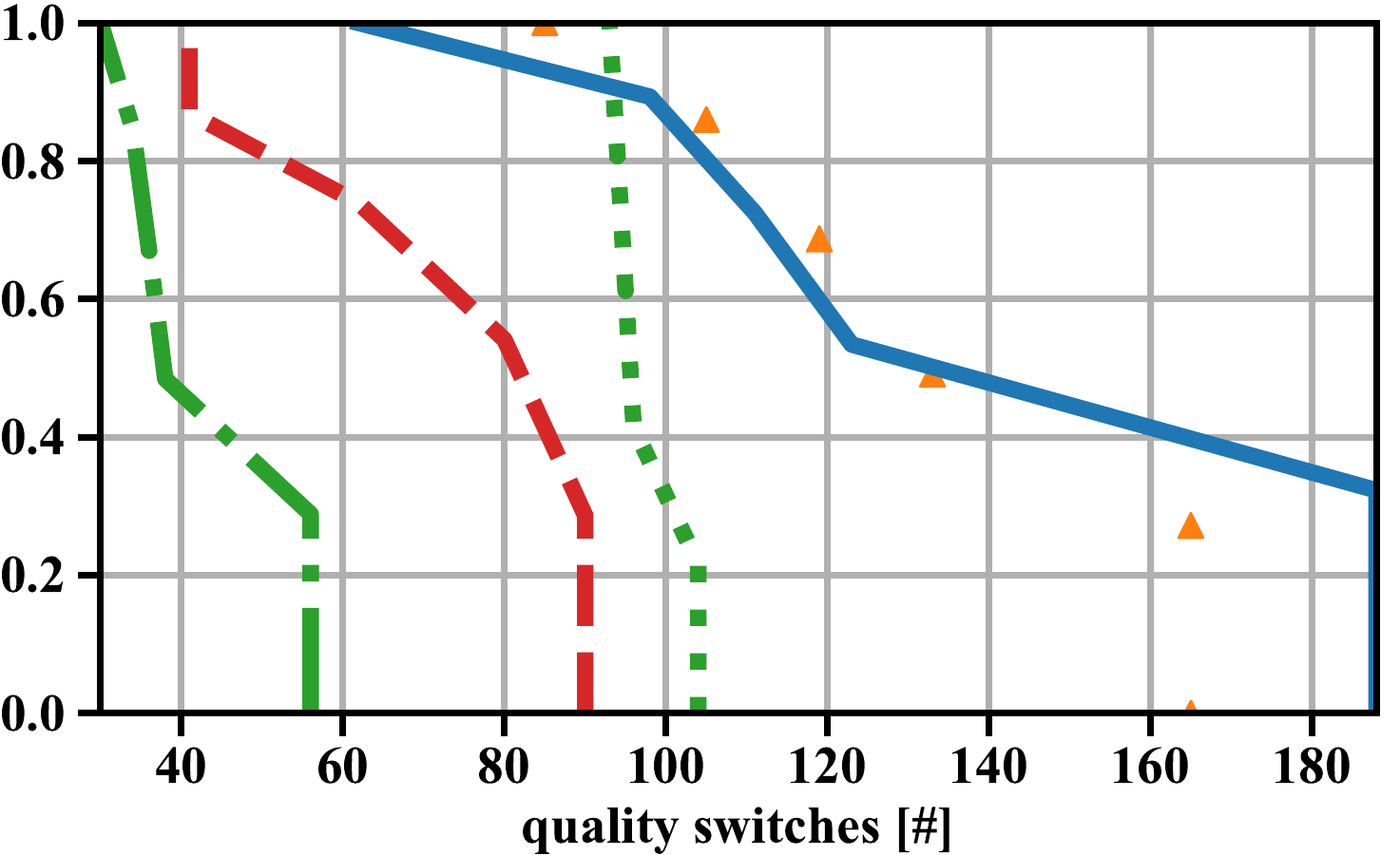}
			\caption{CCDF of the number of quality switches per epoch.}
		\end{subfigure}\hspace*{\fill}
		\begin{subfigure}{0.48\textwidth}
			\includegraphics[width=\linewidth, height=4.6cm]{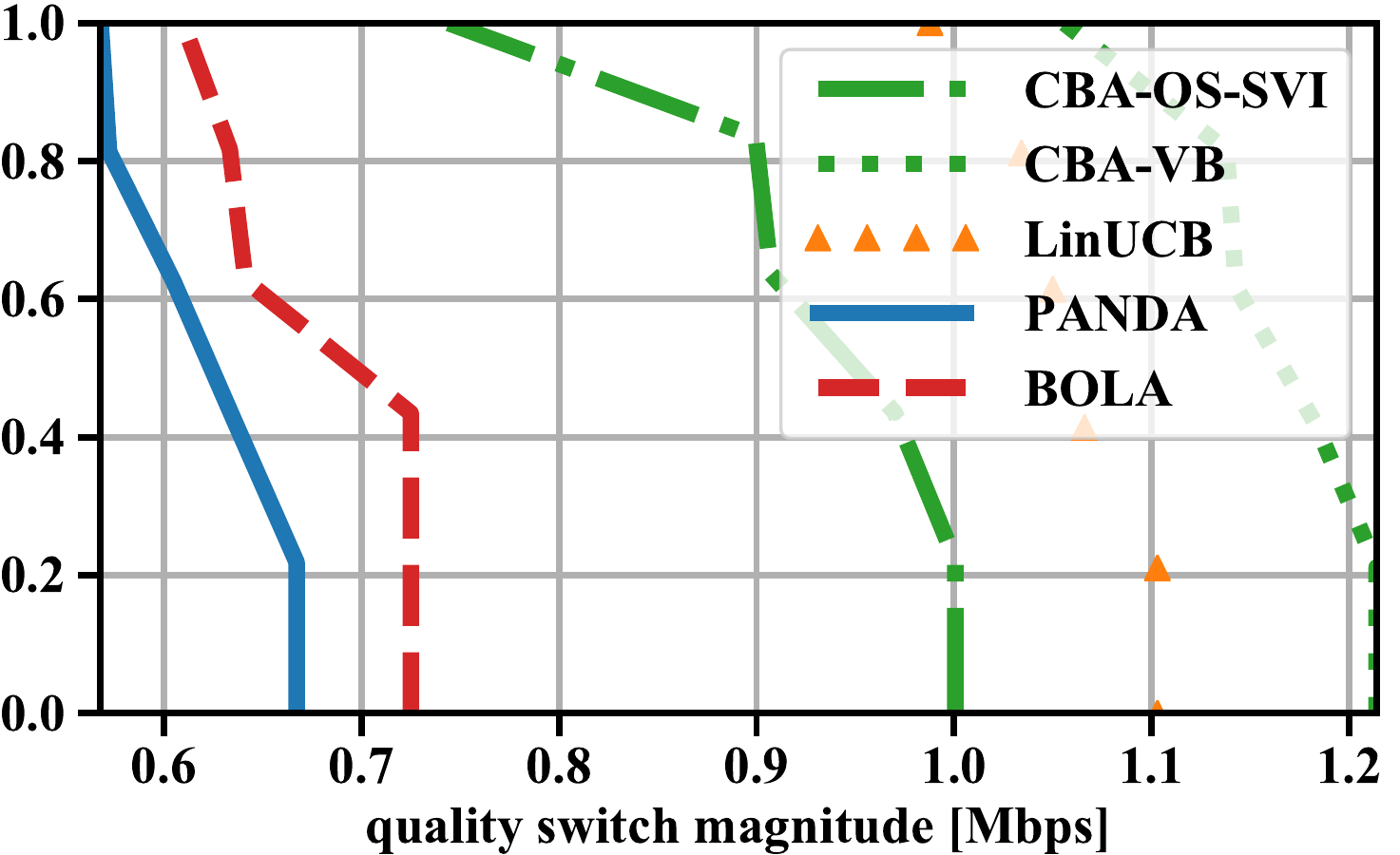}
			\caption{CCDF of the average magnitude of quality switches per epoch.}
		\end{subfigure}
		\caption{Results for full topology with \texttt{HIGH\_QUALITY\_WEIGHTS}}
		\label{fig:full_hq}
	\end{figure*}
	
	%
	
	\begin{figure*}
		\centering
		\begin{subfigure}{0.48\textwidth}
			\includegraphics[width=\linewidth, height=4.6cm]{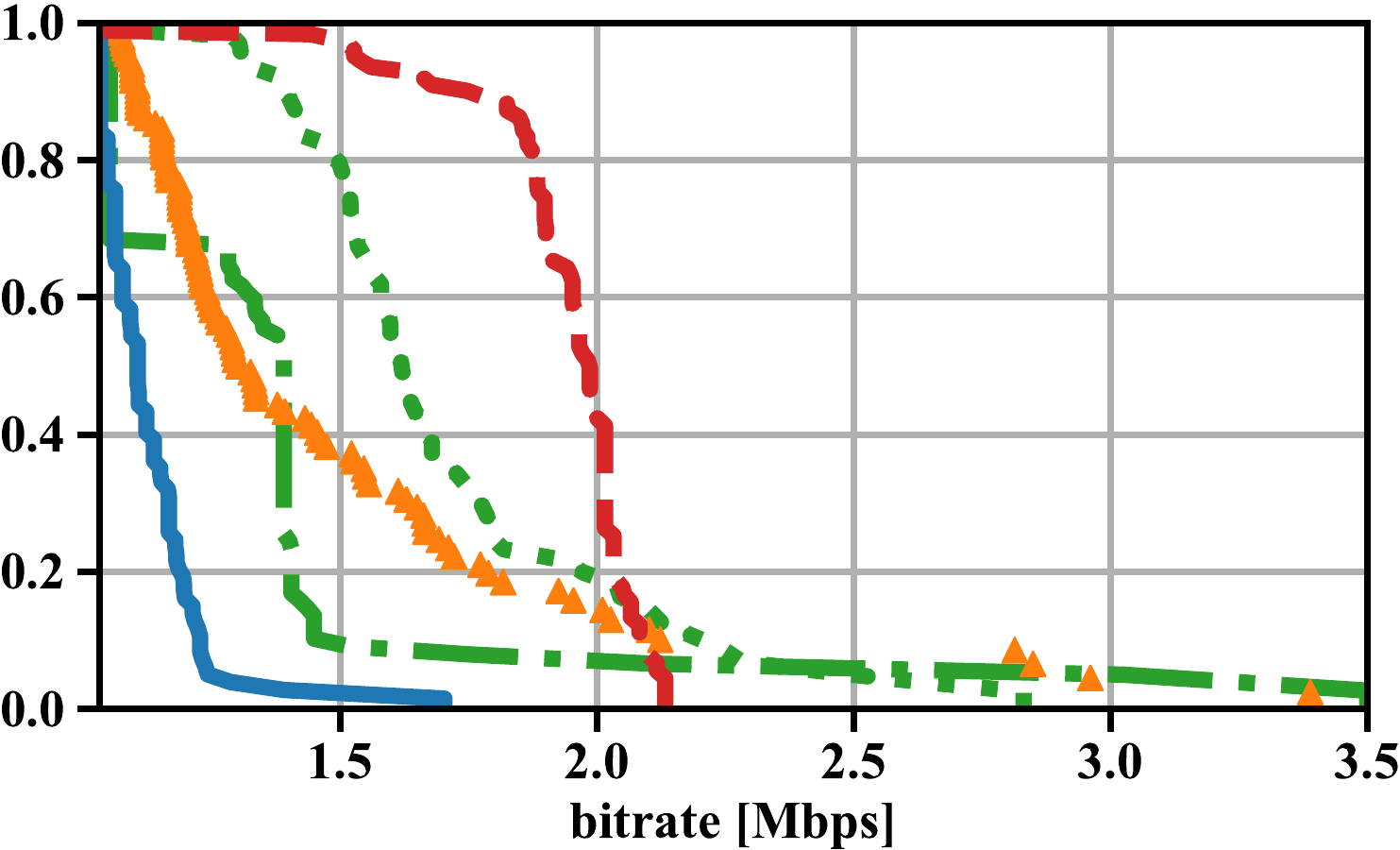}
			\caption{CCDF of average bitrate chosen per epoch.}
		\end{subfigure}\hspace*{\fill}
		\begin{subfigure}{0.48\textwidth}
			\includegraphics[width=\linewidth, height=4.6cm]{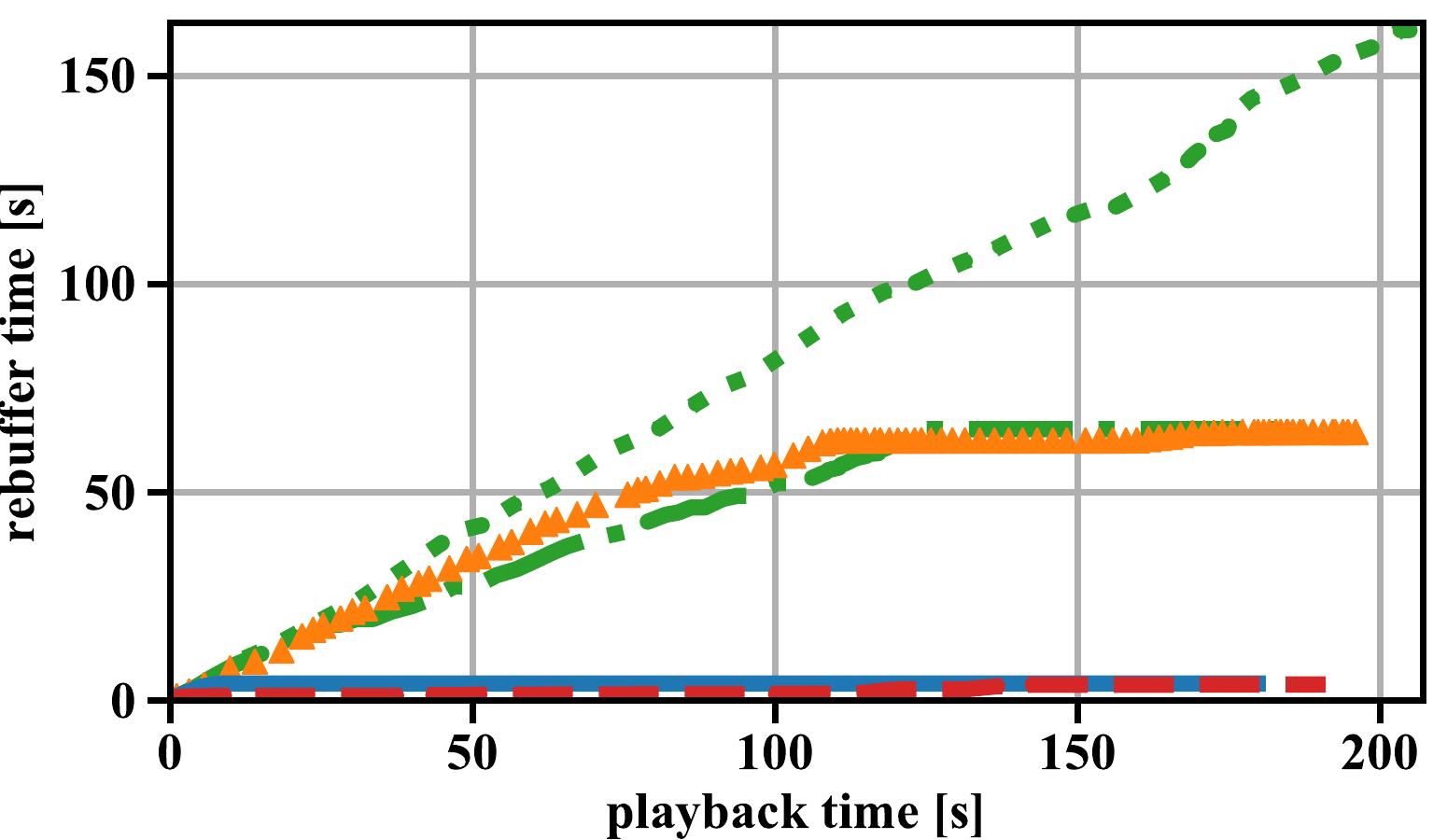}
			\caption{Average cumulative rebuffer time during playback.}
		\end{subfigure}
		
		\medskip
		\begin{subfigure}{0.48\textwidth}
			\includegraphics[width=\linewidth, height=4.6cm]{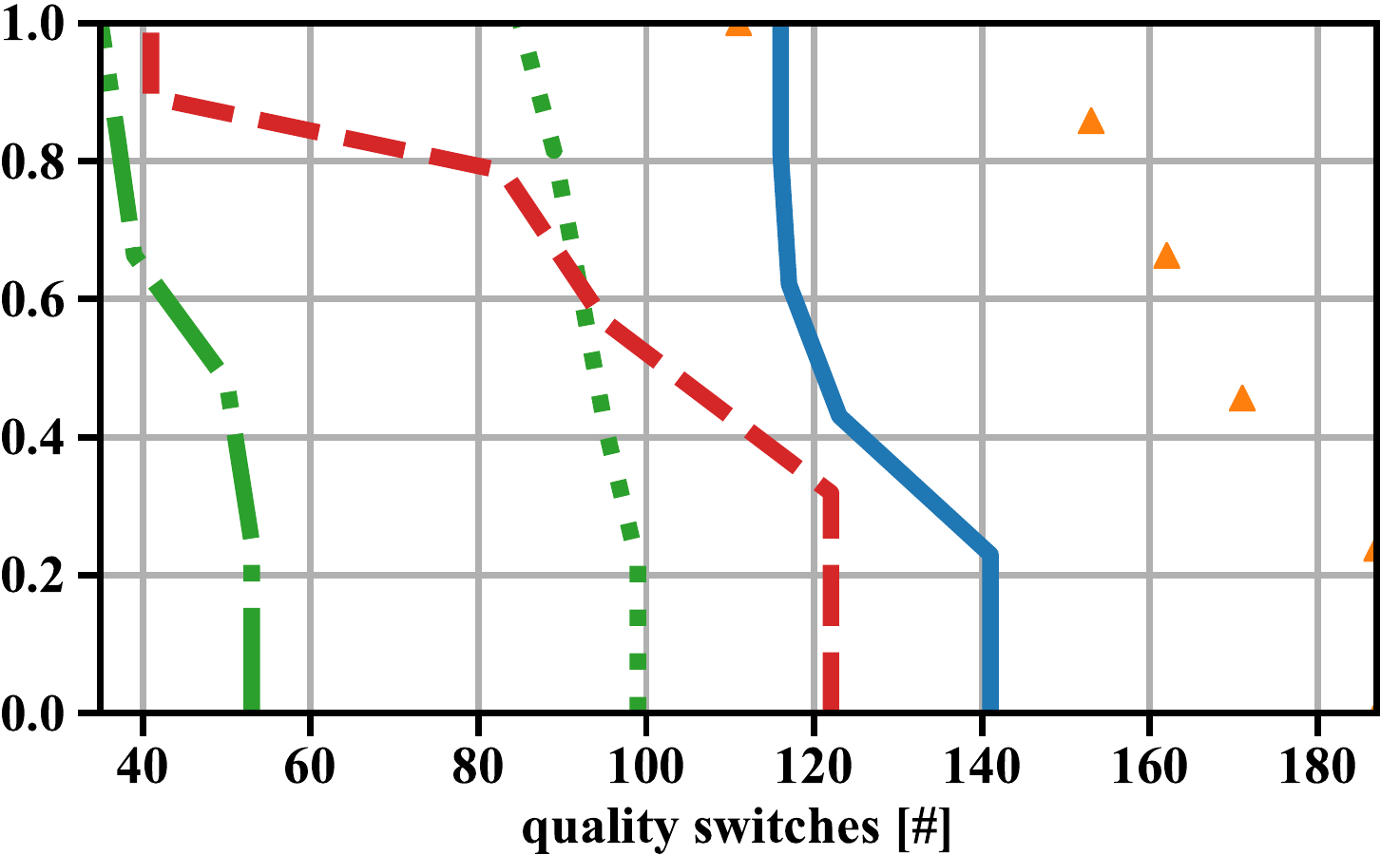}
			\caption{CCDF of the number of quality switches per epoch.}
		\end{subfigure}\hspace*{\fill}
		\begin{subfigure}{0.48\textwidth}
			\includegraphics[width=\linewidth, height=4.6cm]{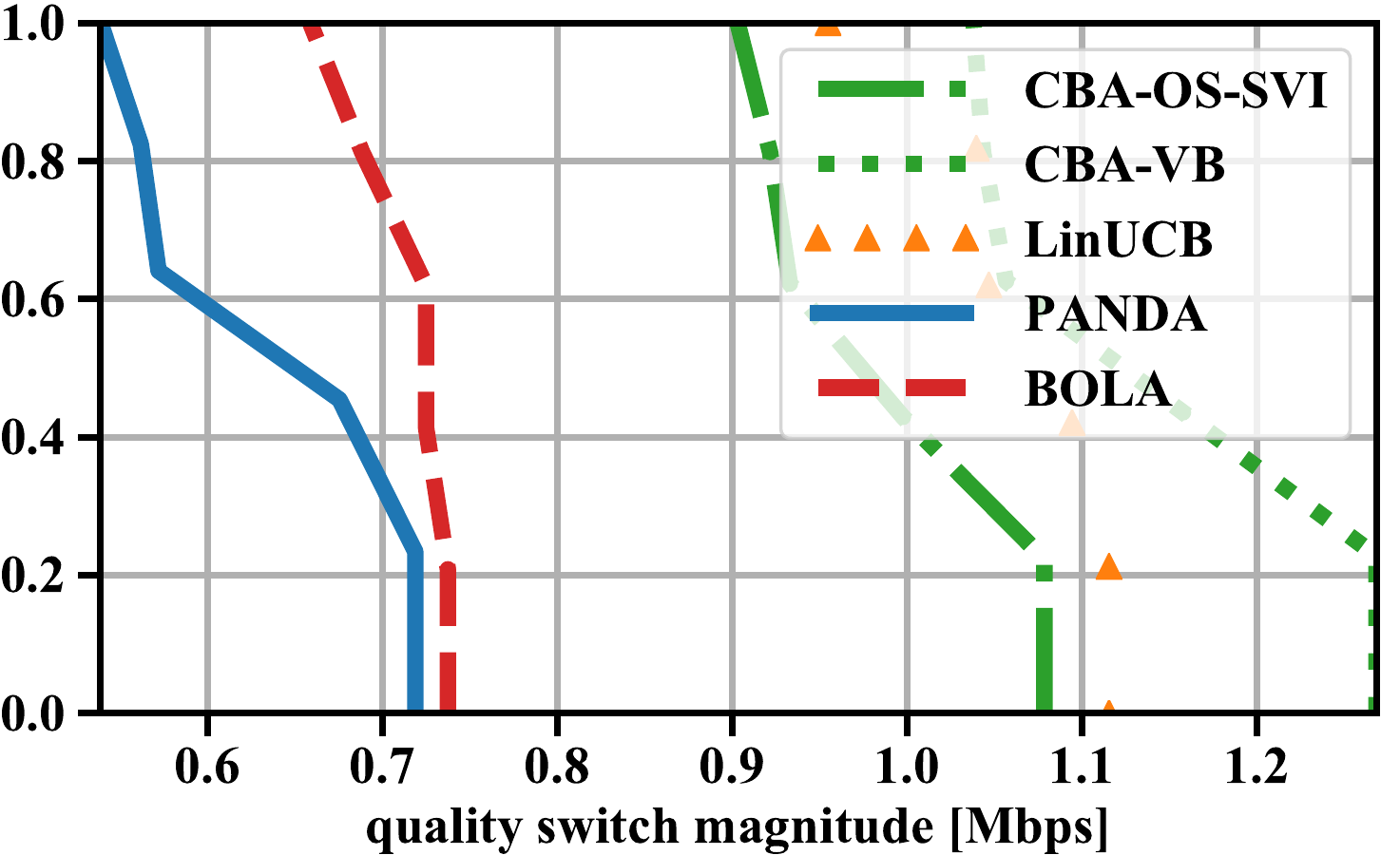}
			\caption{CCDF of the average magnitude of quality switches per epoch.}
		\end{subfigure}
		\caption{Results for the full topology with \texttt{NO\_REBUFFERING\_WEIGHTS}}
		\label{fig:full_nr}
	\end{figure*}
}

\section{Conclusions and Future Work}\label{sec:conclusion}
In this paper, we contributed a sparse Bayesian contextual bandit algorithm for quality adaptation in adaptive video streaming, denoted CBA.
In contrast to state-of-the-art adaptation algorithms, we take high-dimensional video streaming context information and enforce sparsity to shrink the impact of unimportant features.
In this setting, streaming context information includes client-measured variables, such as throughput and buffer filling, as well as, network assistance information.
Since sparse Bayesian estimation is computationally expensive, we developed a fast new inference scheme to support online video quality adaptation.
Furthermore, the provided algorithm is naturally applicable to different adaptive video streaming settings such as DASH over NDN.
Finally, we provided NDN emulation results
showing that CBA yields higher QoE and better QoE fairness between simultaneous streaming sessions compared to
throughput- and buffer-based video quality adaptation algorithms.

\IfShortVersion{\appendices
\section{}
\label{sec:appendix}
\subsection{Parameters of the variational distributions}
\noindent
$\pmb{\mu}_{\pmb{\beta}}=(\mathbf{X}^\top \mathbf{X}+ \mathbf{T}^{-1})^{-1} \mathbf{X}^\top \mathbf{r};\;
\pmb{\Sigma}_{\pmb{\beta}}=\langle \sigma^{-2}\rangle^{-1}  (\mathbf{X}^\top \mathbf{X}+ \mathbf{T}^{-1})^{-1};\;\\
\mathbf{T}^{-1}=\diag(\langle \tau_1^{-1}\rangle,\dots,\langle \tau_D^{-1}\rangle);\;
c^{\ast}=(M+D+c_0)/2;\;\\
d^{\ast}=(\mathbf{r}^\top \mathbf{r} - 2\mathbf{r}^\top\mathbf{X}\langle \pmb{\beta}\rangle + \sum_{m=1}^M \mathbf{x}_m^\top \langle \pmb{\beta}\pmb{\beta}^{\top}\rangle \mathbf{x}_m +\sum_{j=1}^D \langle \beta_j^2 \rangle \langle \tau_j^{-1} \rangle+d_0)/2;\;
p_{\tau, j}=a_0-1/2;\;
a_{\tau, j}=2 \langle \lambda_j \rangle;\;
b_{\tau, j}=\langle \beta_j^2 \rangle \langle \sigma^{-2} \rangle;\;\\
b_{\lambda,j}= \langle \tau_j \rangle + \langle \phi \rangle;\;
a_{\phi}=Db_0+1/2;\;
a_{\lambda,j}=a_0+b_0;\;\\
b_{\phi}= \langle \omega \rangle + \sum_{j=1}^D \langle \lambda_j \rangle;\;
b_{\omega}= \langle \phi \rangle +1;\;
a_{\omega}=1
$

\subsection{Moments of the variational distributions}
\noindent
$\langle \pmb{\beta}\rangle=\pmb{\mu}_{\pmb{\beta}};\;
\langle \pmb{\beta} \pmb{\beta}^\top\rangle= \pmb{\Sigma}_{\pmb{\beta}}+\pmb{\mu}_{\pmb{\beta}}\pmb{\mu}_{\pmb{\beta}}^\top;\;
\langle \sigma^{-2} \rangle = c^\ast / d^\ast;\;\\
\langle \lambda_j \rangle= a_{\lambda,j}/b_{\lambda,j};\;
\langle \phi \rangle = a_{\phi}/ b_{\phi};\;
\langle \omega \rangle = a_{\omega}/b_{\omega};\;\\
v_j=\sqrt{2 a_{\tau, j} b_{\tau, j}};\;
\langle \tau_j \rangle = (b_{\tau, j}/a_{\tau, j})^{1/2} \mathrm{K}(p_{\tau, j}+1, v_j)/\mathrm{K}(p_{\tau, j}, v_j);\;\\
\langle \tau_j^{-1} \rangle =(a_{\tau, j}/b_{\tau, j})^{1/2} \mathrm{K}(1-p_{\tau, j}, v_j)/\mathrm{K}(-p_{\tau, j},v_j)
$

\vspace{5pt}
\noindent $K(p,\cdot)$ is the modified Bessel function of second kind.

\subsection{Intermediate estimates of the natural parameters}
\noindent
$\hat{\pmb{\eta}}_{\pmb{\beta}}=(\langle\sigma^{-2}\rangle M \mathbf{x}_m r_m, -\frac{1}{2} \langle \sigma^{-2}\rangle (M \mathbf{x}_m \mathbf{x}_m^\top+\mathbf{T}^{-1}));\; \\
\hat{\pmb{\eta}}_{\sigma^{-2}}=(\frac{M+D+c_0}{2}-1, -(M r_m^2 - 2M r_m \mathbf{x}_m^\top \langle \pmb{\beta}\rangle + M \mathbf{x}_m^\top \langle \pmb{\beta}\pmb{\beta}^{\top}\rangle \mathbf{x}_m  +\sum_{j=1}^D \langle \beta_j^2 \rangle \langle \tau_j^{-1} \rangle+d_0)/2);\;
\hat{\pmb{\eta}}_{\tau, j}=(a_0-\frac{3}{2}, -\langle \lambda_j \rangle,
\langle \beta_j^2 \rangle \langle\sigma^{-2}\rangle/2);\;\\
\hat{\pmb{\eta}}_{\lambda,j}=(a_0+b_0-1, -\langle\tau_j\rangle-\langle\phi\rangle);\;
\hat{\pmb{\eta}}_{\phi}=(D b_0 -\frac{1}{2},-\langle\omega\rangle -\sum_{j=1}^D \langle\lambda_j\rangle);\;
\hat{\pmb{\eta}}_{\omega}=(0,-\langle\phi\rangle-1)
$

\vspace{5pt}
\noindent The transformation of the natural parametrization is given by

\vspace{5pt}
$(\pmb{\mu}_{\pmb{\beta}}, \pmb{\Sigma}_{\pmb{\beta}})=(-\frac{1}{2}{{\pmb{\eta}^{(2)}_{\pmb{\beta}}}}^{-1}{\pmb{\eta}^{(1)}_{\pmb{\beta}}},-\frac{1}{2} {{\pmb{\eta}^{(2)}_{\pmb{\beta}}}}^{-1});\;\\
(c^\ast,d^\ast)=( {{{\eta}^{(1)}_{\sigma^{-2}}}}+1,- {{{\eta}^{(2)}_{\sigma^{-2}}}});\;
(p_{\tau, j}, a_{\tau, j}, b_{\tau, j})=({{\eta}^{(1)}_{\tau, j}}+1,-2{{\eta}^{(2)}_{\tau, j}}, 2{{\eta}^{(3)}_{\tau, j}});\;
(a_{\lambda,j}, b_{\lambda,j})=({\eta^{(1)}_{\lambda,j}}+1, -{\eta^{(2)}_{\lambda,j}});\;
(a_{\phi}, b_{\phi})=({\eta^{(1)}_{\phi}}+1, -{\eta^{(2)}_{\phi}});\;
(a_{\omega}, b_{\omega})=({\eta^{(1)}_{\omega}}+1, -{\eta^{(2)}_{\omega}})
$

\vspace{5pt}
\noindent $\eta^{(i)}$ is the $i$-th variable of the tuple of natural parameters $\pmb{\eta}$.
}
\IfLongVersion{\appendices
\section{The Generalized Inverse Gaussian}
\label{sec:appendix_gig}
The probability density function of a generalized inverse Gaussian (GIG) distribution is
\begin{equation}
\begin{split}
&\GIGDis(x \mid p, a, b)= (a/b)^{p/2}(2 \mathrm{K}_{p}(\sqrt{ab}))^{-1} \\
&\qquad\qquad\qquad \qquad\qquad x^{p-1} \exp\{(ax+b/x)/2\}\\
\end{split}
\end{equation}
The GIG distribution with parameters $\pmb{\theta}=[p,a,b]^\top$ is a member of the exponential family distribution with base measure $h(x)=1$, natural parameters $\pmb{\eta}(\pmb{\theta})=[p-1,-a/2,b/2]^\top$, sufficient statistics $\mathbf{S}(x)=[log(x), x, 1/x]^\top$ and log-normalizer $A(\pmb{\eta})= \log((-\eta^{(2)}/\eta^{(3)})^{(\eta^{(1)}+1)/2} \{2 \mathrm{K}_{\eta^{(1)}+1}(\sqrt{-4\eta^{(2)}\eta^{(3)}})\}^{-1})$. The inverse transform of the natural parameters is obtained by $\pmb{\theta}(\pmb{\eta})=[\eta^{(1)}+1,-2\eta^{(2)},2\eta^{(3)}]^\top$.

\section{Calculation of the ELBO}
\label{sec:appendix_elbo}
Here, we present the calculation for the ELBO. The joint distributions involved in the calculation of the evidence lower bound \eqref{eq:elbo_general} factorize as
\begin{equation}
\begin{split}
&p(\pmb{\beta}, \sigma^{-2}, \pmb{\tau},\pmb{\lambda}, \phi, \omega , \mathbf{r})=p(\mathbf{r} \mid \pmb{\beta}, \sigma^{-2}) p(\pmb{\beta}\mid \sigma^{-2}, \pmb{\tau} ) \\
&\qquad p(\sigma^{-2})p(\pmb{\tau}\mid \pmb{ \lambda})p( \pmb{ \lambda} \mid \phi) p(\phi\mid\omega)p(\omega)
\end{split}
\label{eq:p_factorization}
\end{equation}
and
\begin{equation}
q(\pmb{\beta}, \sigma^{-2}, \pmb{\tau},\pmb{\lambda}, \phi, \omega)=q(\pmb{\beta})q(\sigma^{-2})q( \pmb{\tau})q(\pmb{\lambda})q(\phi)q(\omega)
\label{eq:q_factorization}
\end{equation}
Denoting $\langle \cdot \rangle$ as the expactation w.r.t. to the distribution $q$, the evidence lower bound  \eqref{eq:elbo_general} is
\begin{equation}
\begin{split}
&\mathcal{L}(q)=\langle \log p(\mathbf{r} \mid \pmb{\beta}, \sigma^{-2}) \rangle
+\langle \log p(\pmb{\beta}\mid \sigma^{-2}, \pmb{\tau} ) \rangle \\
&+\langle \log p(\sigma^{-2})\rangle
+\langle \log p(\pmb{\tau}\mid \pmb{ \lambda})\rangle
+\langle \log p( \pmb{ \lambda} \mid \phi) \rangle\\
&+\langle \log p(\phi\mid\omega)\rangle
 + \langle \log p(\omega)\rangle
-\langle \log q(\pmb{\beta})\rangle\\
&-\langle \log q(\sigma^{-2})\rangle
-\langle \log q( \pmb{\tau})\rangle
-\langle \log q(\pmb{\lambda})\rangle\\
&-\langle \log q(\phi)\rangle
-\langle \log q(\omega)\rangle\\
\end{split}
\label{eq:elbo_factorized}
\end{equation}
The expected values of the log factorized joint distribution \eqref{eq:p_factorization} needed for \eqref{eq:elbo_factorized} are
\begin{equation}
\begin{split}
&\langle \log(p(\mathbf{r}\mid \pmb{\beta}, \sigma^{-2})) \rangle = -M/2 \log(2\pi) + M/2 \langle \log(\sigma^{-2}) \rangle \\
&\quad- 1/2 \langle \sigma^{-2} \rangle (\mathbf{r}^\top \mathbf{r} -2\mathbf{r}^\top \langle\pmb{\beta}\rangle + \sum_{m=1}^M \mathbf{x}_m^\top \langle\pmb{\beta} \pmb{\beta}^\top \rangle \mathbf{x}_m)\\
&\langle \log(p(\pmb{\beta}\mid \sigma^{-2},\pmb{\tau})) \rangle = -D/2 \log(2\pi) + D/2 \langle \log(\sigma^{-2}) \rangle \\
&\quad-1/2 \sum_{j=1}^D \langle\log(\tau_j)\rangle -1/2 \sum_{j=1}^D \langle \sigma^{-2} \rangle \langle \beta_j^2 \rangle \langle \tau_j^{-1}\rangle\\
&\langle \log(p(\sigma^{-2}))\rangle=(c_0/2-1) \langle \log(\sigma^{-2}) \rangle -d_0/2 \langle \sigma^{-2} \rangle \\
&\quad+c_0/2 \log(d_0/2)- log(\Gamma(c_0/2))\\
&\langle \log(p(\pmb{\tau}\mid \pmb{\lambda})) \rangle = - D \log(\Gamma(a_0))+ a_0 \sum_{j=1}^D \langle\log(\lambda_j)\rangle \\
&\quad+ (a_0-1) \sum_{j=1}^D  \langle \log(\tau_j)\rangle - \sum_{j=1}^D \langle\lambda_j\rangle \langle\tau_j\rangle\\
&\langle \log( p(\pmb{\lambda}\mid \phi)) \rangle = - D \log(\Gamma(b_0)) + b_0 D \langle \log(\phi) \rangle \\
&\quad+ (b_0-1)\sum_{j=1}^D \langle \log(\lambda_j) \rangle - \langle \phi \rangle \sum_{j=1}^D \langle \lambda_j \rangle\\
& \langle \log( p(\phi \mid \omega)) \rangle = \frac{1}{2} \langle \log(\omega)\rangle  - \log(\Gamma(\frac{1}{2})) - \frac{1}{2}\langle \log(\phi) \rangle  \\
&\quad- \langle \omega \rangle \langle \phi \rangle\\
& \langle \log(p(\omega)) \rangle = \frac{1}{2} \log(\frac{1}{2}) - \log(\Gamma(\frac{1}{2})) -\frac{1}{2} \langle \log(\omega)\rangle \\
&\quad-\frac{1}{2} \langle \omega \rangle
\end{split}
\end{equation}
The expected values of the log factorized variational distribution \eqref{eq:q_factorization} compute to
\begin{equation}
\begin{split}
&\langle \log(q(\pmb{\beta})) \rangle=- \frac{D}{2} \log(2\pi)- \frac{1}{2} \log(\vert \pmb{\Sigma}_{\pmb{\beta}} \vert)\\
&\langle \log(q( \sigma^{-2})) \rangle= c^\ast \log(d^\ast)-\log(\Gamma(c^\ast))\\
&\qquad+ (c^\ast-1) \langle \log(\sigma^{-2}) \rangle -d^\ast \langle \sigma^{-2} \rangle\\
& \langle \log(q(\pmb{\tau})) \rangle = (\frac{a_0}{2}-\frac{1}{2}) \left[p\log(2)-p\log(\langle\sigma^{-2}\rangle)\right.\\
&\quad \left.+\sum_{j=1}^D\log(\langle\lambda_j\rangle) -\sum_{j=1}^D \log(\langle\beta_j^2\rangle)\right]-p\log(2)\\
&\quad -\sum_{j=1}^D \log(\mathrm{K}_{a_0-\frac{1}{2}}(v_j))+(a_0-\frac{3}{2})\sum_{j=1}^D \langle\log(\tau_j)\rangle \\
&\qquad-\sum_{j=1}^D \langle \lambda_j\rangle \langle \tau_j\rangle - \frac{1}{2} \langle \sigma^{-2} \rangle \sum_{j=1}^D \langle \beta_j^2 \rangle \langle \tau_j^{-1}\rangle\\
&\langle \log(q(\pmb{\lambda})) \rangle=-D\log(\Gamma(a_0+b_0)) +(a_0+b_0)\\
&\quad \times \sum_{j=1}^D \log(\langle\tau_j\rangle+\langle\phi\rangle)+(a_0+b_0-1)\sum_{j=1}^D \langle\log(\lambda_j)\rangle \\
&\quad- \sum_{j=1}^D \langle\tau_j\rangle\langle\lambda_j\rangle-\langle\phi\rangle\sum_{j=1}^D\langle\tau_j\rangle \\
\end{split}
\end{equation}
\begin{equation*}
\begin{split}
& \langle \log(q(\phi)) \rangle=(D b_0+\frac{1}{2})\log(\langle\omega \rangle + \sum_{j=1}^D \langle \lambda_j \rangle)\\
&\qquad- \log(\Gamma(D b_0+\frac{1}{2}))+(D b_0-\frac{1}{2}) \langle \log(\phi) \rangle  \\
&\qquad-\langle\omega\rangle\langle\phi\rangle- \langle\phi\rangle\sum_{j=1}^D\langle\lambda_j\rangle   \\
&\langle \log(q(\omega)) \rangle=\log(\langle \phi \rangle +1) -\langle \omega \rangle\langle \phi \rangle -\langle \omega \rangle
\end{split}
\end{equation*}
Therefore, we can calculate the evidence lower bound \eqref{eq:elbo_general} as
\begin{equation}
\begin{split}
&\mathcal{L}=-\frac{M}{2} \log(2\pi) +\frac{c_0}{2}\log(\frac{d_0}{2})-\log(\Gamma(\frac{c_0}{2}))\\
&-c^\ast\log(d^\ast) + log(\Gamma(c^\ast))+\sum_{j=1}^D \langle \lambda_j\rangle\langle\tau_j\rangle+\langle\phi\rangle\sum_{j=1}^D \langle\lambda_j\rangle  \\
&- D\log(\Gamma(a_0)) - D\log(\Gamma(b_0)) -2 \log(\Gamma(\frac{1}{2}))\\
& +\log(2)((\frac{5}{4}-\frac{a_0}{2})D-\frac{1}{2})+D\log(\Gamma(a_0+b_0)) \\
&+\log(\Gamma(p D+\frac{1}{2}))+\frac{1}{2}\langle\omega\rangle+\langle\omega\rangle\langle\phi\rangle+\frac{1}{2}\log(\vert \pmb{\Sigma}_{\pmb{\beta}} \vert)\\
\end{split}
\end{equation}
\begin{equation*}
\begin{split}
&+(\frac{a_0}{2}-\frac{1}{4})(D\log(\langle\sigma^{-2}\rangle)-\sum_{j=1}^D \log(\langle\lambda_j\rangle)) \\
 &+(\frac{a_0}{2}-\frac{1}{4})\sum_{j=1}^D \log(\langle\beta_j^2\rangle)-\sum_{j=1}^D \log(\mathrm{K}_{a_0-\frac{1}{2}}(v_j))\\
 &-\frac{1}{2}\langle\sigma^{-2}\rangle \sum_{j=1}^D \langle\beta_j^2\rangle\langle \tau_j^{-1}\rangle -(D b_0+\frac{1}{2})\log(\langle\omega \rangle\\
 & +\sum_{j=1}^D\langle\lambda_j\rangle)-(a_0+b_0)\sum_{j=1}^D \log(\langle\tau_j\rangle+\langle\phi\rangle)-\log(\langle\phi\rangle+1).
\end{split}
\end{equation*}
\section{Calculation for the intermediate estimates of the natural parameters}
\label{sec:appendix_intermediate}
In order to calculate \eqref{eq:stoch_gradient_update} we need to calculate the intermediate parameters $\hat{\pmb{\eta}}$ for all parameters, \ie $\pmb{\beta}$ , $\sigma^{-2}$, $\pmb{\tau}$, $\pmb{\lambda}$, $\phi$ and $\omega$. For this we calculate $\tilde{\pmb{\eta}}=\mathrm{E}_q[\pmb{\eta}^{\prime}]$, where $\pmb{\eta}^{\prime}$ is the natural parameter of the full conditional for the corresponding variational factor.

Therefore, we compute the full conditionals for the parameters
\begin{equation}
\begin{split}
&p(\pmb{\beta} \mid \mathbf{r}, \mathbf{X}, \sigma^{-2}, \pmb{\tau})=\NormalDis(\pmb{\beta} \mid \pmb{\mu}_{\pmb{\beta}}^{\prime}, \pmb{\Sigma}_{\pmb{\beta}}^{\prime})\\
   &p( \sigma^{-2} \mid \mathbf{r}, \mathbf{X}, \pmb{\beta}, \pmb{\tau} )= \GammaDis(\sigma^{-2} \mid c^\prime, d^\prime)\\
   &p(\pmb{\tau} \mid \pmb{\beta}, \sigma^{-2}, \pmb{\lambda} ) = \prod_{j=1}^{D} \GIGDis(\tau_j  \mid p_{\tau,j}^{\prime}, a_{\tau,j}^{\prime}, b_{\tau,j}^{\prime})\\
   &p(\pmb{\lambda} \mid \pmb{\tau}, \phi )=\prod_{j=1}^{D} \GammaDis(\lambda_j \mid a_{\lambda,j}^{\prime},b_{\lambda,j}^{\prime})\\
   &p(\phi \mid \pmb{\lambda}, \omega)= \GammaDis(\phi \mid a_{\phi}^{\prime}, b_{\phi}^{\prime}) \\
   &p(\omega \mid \phi)= \GammaDis(\omega \mid a_{\omega}^{\prime},b_{\omega}^{\prime}),
   \end{split}
\end{equation}
with the parameters
\begin{equation}
\begin{split}
&\pmb{\mu}_{\pmb{\beta}}^{\prime}=(\mathbf{X}^\top \mathbf{X}+ {\mathbf{T}^{\prime}}^{-1})^{-1} \mathbf{X}^\top \mathbf{r},\\
  & \pmb{\Sigma}_{\pmb{\beta}}^{\prime}=1/\sigma^{-2} (\mathbf{X}^\top \mathbf{X}+ {\mathbf{T}^{\prime}}^{-1})^{-1},\\
  & {\mathbf{T}^{\prime}}^{-1}=\diag(\tau_1^{-1},\dots,\tau_D^{-1}),\\
    &c^{\prime}=\frac{M+D+c_0}{2},\\
   &d^{\prime}=\frac{(\mathbf{r}-\mathbf{X}\pmb{\beta})^\top(\mathbf{r}-\mathbf{X}\pmb{\beta})+\pmb{\beta}^\top {\mathbf{T}^{\prime}}^{-1} \pmb{\beta}+d_0}{2},\\
   &p_{\tau,j}^{\prime}=a_0-1/2,\quad
   a_{\tau,j}^{\prime}=2 \lambda_j, \quad
   b_{\tau,j}^{\prime}= \beta_j^2\sigma^{-2},\\
   &a_{\lambda,j}^{\prime}=a_0+b_0, \quad
   b_{\lambda,j}^{\prime}=\tau_j + \phi \\
   &a_{\phi}^{\prime}=D b_0+1/2, \quad
    b_{\phi}^{\prime}=\sum_{j=1}^D \lambda_j +\omega\\
    &a_{\omega}^{\prime}=1, \quad
    b_{\omega}^{\prime}=\phi+1
\end{split}
\end{equation}
Next, we transform the parameters of the full conditionals into the exponential family parametrization.

The inverse transform of the natural parameters of the full conditionals is given by
\begin{equation}
\begin{split}
&\pmb{\eta}_{\pmb{\beta}}^{\prime}=\left({\pmb{\Sigma}_{\pmb{\beta}}^{\prime}}^{-1}\pmb{\mu}_{\pmb{\beta}}^{\prime},-\frac{1}{2}{\pmb{\Sigma}_{\pmb{\beta}}^{\prime}}^{-1}\right)\\
&{\pmb{\eta}^{\prime}_{\sigma^{-2}}}=\left(c^{\prime}-1, -d^{\prime}\right)\\
&{\pmb{\eta}^{\prime}_{\tau,j}}=\left(p_{\tau,j}^{\prime}-1, a_{\tau,j}^{\prime}/2, b_{\tau,j}^{\prime}/2\right)\\
&\pmb{\eta}_{\lambda,j}^{\prime}=\left(a_{\lambda,j}^{\prime}-1, -b_{\lambda,j}^{\prime}\right)\\
&\pmb{\eta}_{\phi}^{\prime}=\left(a_{\phi}^{\prime}-1, -b_{\phi}^{\prime}\right)\\
&\pmb{\eta}_{\omega}^{\prime}=\left(a_{\omega}^{\prime}-1, -b_{\omega}^{\prime}\right).
\end{split}
\end{equation}

Taking the expected value w.r.t. the variational distribution $q$ we calculate the parameters $\tilde{\pmb{\eta}}=\mathrm{E}_q[\pmb{\eta}^{\prime}]$ as
\begin{equation}
\begin{split}
&\tilde{\pmb{\eta}}_{\pmb{\beta}}=\left(\langle\sigma^{-2}\rangle \mathbf{X}^\top \mathbf{r}, -\frac{1}{2} \langle \sigma^{-2}\rangle (\mathbf{X}^\top \mathbf{X}+\mathbf{T}^{-1})\right),\;  \\
&\tilde{\pmb{\eta}}_{\sigma^{-2}}=\left(\frac{M+D+c_0}{2}-1, -(\mathbf{r}^\top\mathbf{r} - 2\mathbf{r}^\top\mathbf{X}  \langle \pmb{\beta}\rangle\right.\\
&\qquad\quad + \left.\sum_{m=1}^M \mathbf{x}_m^\top \langle \pmb{\beta}\pmb{\beta}^{\top}\rangle \mathbf{x}_m  +\sum_{j=1}^D \langle \beta_j^2 \rangle \langle \tau_j^{-1} \rangle+d_0)/2\right),\;  \\
&\tilde{\pmb{\eta}}_{\tau,j}=\left(a_0-\frac{3}{2}, -\langle \lambda_j \rangle, \langle \beta_j^2 \rangle \langle\sigma^{-2}\rangle/2\right),\;\\
&\tilde{\pmb{\eta}}_{\lambda,j}=\left(a_0+b_0-1, -\langle\tau_j\rangle-\langle\phi\rangle\right),\;\\
& \tilde{\pmb{\eta}}_{\phi}=\left(D b_0 -\frac{1}{2},-\langle\omega\rangle -\sum_{j=1}^D \langle\lambda_j\rangle\right),\;\\
&\tilde{\pmb{\eta}}_{\omega}=\left(0,-\langle\phi\rangle-1\right).
\end{split}
\end{equation}

Replicating one data point $(\mathbf{x}_m, r_m)$ $M$ times yields the intermediate estimates
\begin{equation}
\begin{split}
&\hat{\pmb{\eta}}_{\pmb{\beta}}=\left(\langle\sigma^{-2}\rangle M \mathbf{x}_m r_m, -\frac{1}{2} \langle \sigma^{-2}\rangle (M \mathbf{x}_m \mathbf{x}_m^\top+\mathbf{T}^{-1})\right),\;  \\
&\hat{\pmb{\eta}}_{\sigma^{-2}}=\left(\frac{M+D+c_0}{2}-1, -(M r_m^2 - 2M r_m \mathbf{x}_m^\top \langle \pmb{\beta}\rangle\right.\\
&\qquad\qquad + \left.M \mathbf{x}_m^\top \langle \pmb{\beta}\pmb{\beta}^{\top}\rangle \mathbf{x}_m  +\sum_{j=1}^D \langle \beta_j^2 \rangle \langle \tau_j^{-1} \rangle+d_0)/2\right),\;  \\
&\hat{\pmb{\eta}}_{\tau,j}=\left(a_0-\frac{3}{2}, -\langle \lambda_j \rangle, \langle \beta_j^2 \rangle \langle\sigma^{-2}\rangle/2\right),\;\\
&\hat{\pmb{\eta}}_{\lambda,j}=\left(a_0+b_0-1, -\langle\tau_j\rangle-\langle\phi\rangle\right),\;\\
& \hat{\pmb{\eta}}_{\phi}=\left(D b_0 -\frac{1}{2},-\langle\omega\rangle -\sum_{j=1}^D \langle\lambda_j\rangle\right),\;\\
&\hat{\pmb{\eta}}_{\omega}=\left(0,-\langle\phi\rangle-1\right).
\end{split}
\end{equation}

}

\bibliographystyle{IEEEbib}
\balance
\bibliography{bibliography_main}

\end{document}